# Outcome-wide longitudinal designs for causal inference: a new template for empirical studies

Tyler J. VanderWeele
Harvard T.H. Chan School of Public Health
Institute for Quantitative Social Science, Harvard University

Maya B. Mathur
Harvard T.H. Chan School of Public Health
Quantitative Sciences Unit, Stanford University

Ying Chen
Institute for Quantitative Social Science, Harvard University

*Abstract*

In this paper we propose a new template for empirical studies intended to assess causal effects: the outcome-wide longitudinal design. The approach is an extension of what is often done to assess the causal effects of a treatment or exposure using confounding control, but now, over numerous outcomes. We discuss the temporal and confounding control principles for such outcome-wide studies, metrics to evaluate robustness or sensitivity to potential unmeasured confounding for each outcome, and approaches to handle multiple testing. We argue that the outcome-wide longitudinal design has numerous advantages over more traditional studies of single exposure-outcome relationships including results that are less subject to investigator bias, greater potential to report null effects, greater capacity to compare effect sizes, a tremendous gain in the efficiency for the research community, a greater policy relevance, and a more rapid advancement of knowledge. We discuss both the practical and theoretical justification for the outcome-wide longitudinal design and also the pragmatic details of its implementation.

# 1. INTRODUCTION

In much biomedical and social science research intended to assess causal effects with observational data, a particular template or structure to analysis and reporting is frequently employed. When the effect of some treatment or exposure is to be assessed on a particular outcome, it is frequently the case that a regression model is fit for the outcome conditional on the exposure or treatment and a number of covariates. Ideally, the outcome occurs temporally subsequent to the exposure, and the covariate values pertain to a period temporally before the exposure or are at least not affected by the exposure. Confidence intervals, p-values, and other measures of uncertainty are reported for the regression coefficient for the exposure and this is then often interpreted as an estimate of the causal effect of the exposure on the outcome. Formal systems related to potential outcomes or causal diagrams have been developed that justify this approach and interpretation, and clarify under what assumptions it holds (Pearl, 2009; Imbens and Rubin, 2015; Morgan and Winship, 2015; Hernán and Robins, 2018).

There are certainly variations to this basic template. Not infrequently analyses are also stratified by one or more other variables, such as gender or race, to see if the effect estimates vary across groups. Sometimes more sophisticated modeling strategies or machine learning algorithms are used to obtain estimates of the causal effect on the desired effect scale (e.g. van der Laan and Rose, 2011, 2018; Belloni et al, 2014; Schuler and Rose, 2017). Sometimes, albeit not very frequently, sensitivity analysis or bias analysis techniques are used to assess how robust or sensitive conclusions are to the presence of uncontrolled confounding, or measurement error, or selection bias (Rosenbaum and Rubin, 1983a; Rothman et al., 2008; Lash et al., 2009; Ding and VanderWeele, 2016).

Certainly not all studies intended to assess causal effects conform to this template. Some studies address more complex inquiries concerning the effects of time-varying exposures (Robins, 1992, Robins et al., 2000; Robins and Hernán, 2009; Hernán and Robins, 2018) or whether some effects are mediated by others (Imai et al., 2010; VanderWeele, 2015). Others, even when attempting only to assess the effect of an exposure at a single point in time, employ instrumental variables, rather than covariate control, to attempt to address issues of confounding (Angrist et al., 1995). Other approaches especially in econometrics use discontinuities in treatment assignment, or differences in trends, or sudden unpredictable shocks and events to attempt to identify causal effects and various suites of methods often referred to respectively as regression discontinuities designs, difference-in-difference methods, and interrupted time-series designs have been developed to address these settings (Angrist and Pischke, 2009; Morgan and Winship, 2015). Nevertheless, in the biomedical sciences at least, the covariate-controlled regression approach is still perhaps used with the greatest frequency. And with well-designed studies, it has often, though perhaps not always, served the research community reasonably well.

Reasonable criticisms are often still leveled against this template. There is of course always the possibility that unmeasured confounding may still bias effect estimates even when extensive effort has been made to control for as many pre-exposure covariates as possible related to both the exposure and the outcome. This threat of unmeasured confounding is almost always present with observational data. This basic template has also been criticized on the grounds that in practice it allows investigators too many degrees of freedom in the decisions as to how to go about modeling the outcome or what covariates to control for (Simmons et al., 2011; Gelman and Loken, 2014). Investigators may be tempted to fit many different models and choose the ones that best conform to their hopes and expectations. Even those who desire to maintain integrity may end up having to make such choices across models inadvertently. Recent machine learning approaches that use cross validation to make choices across many models may help in part obviate the need for such choices (van der Laan and Rose, 2011, 2018), but are still employed relatively infrequently, and still require the investigator to make decisions on the list of covariates to input into these algorithms, once again introducing investigator choice. The basic template has also been criticized on the grounds of its effect on science, taken cumulatively, over numerous studies. Investigators, reviewers and journal editors not infrequently use a p-value cut-off of 0.05 to assess whether there is evidence for an effect. However, across thousands and millions of studies of investigators across the globe, the cumulative effect of speaking of having "discovered effects" whenever the p-value is below 0.05 is having numerous false positive published in the literature (Head et al. 2015). This in combination with the previous potential biases due to unmeasured confounding and investigator discretion has led some to conclude that perhaps the majority of research findings in the literature are "false" (Ioannidis, 2005). These phenomena are likely also in part responsible for the recent so-called "replication crisis" (Open Science Collaboration, 2015; Camerer et al., 2016).

In this paper we would like to propose a development or expansion of the current template that we believe will help in part address these various criticisms. We will refer to this new basic template, an extension and expansion of the existing one, as "Outcome-wide longitudinal design" for causal inference. The basic idea of this new template is to make use of the existing template for a single exposure but simultaneously apply it to multiple outcomes, temporally subsequent to the exposure, while supplementing these analyses with new metrics to address potential unmeasured confounding and multiple testing. We propose to address the prior criticisms of the existing template in a number of ways. First, we propose to address the potential bias due to unmeasured confounding by always reporting a new metric, called the E-value (VanderWeele and Ding, 2017), related to how sensitive or robust estimates are to one or more potential unmeasured confounders. Second, we propose that in these outcome-wide studies, decisions about covariate control, and about basic forms of modeling, be made for all outcomes simultaneously according to principles laid out below. The simultaneous decisions for all outcomes, while not eliminating investigator discretion entirely, does limit it substantially because if decisions are made to "optimize" the results for one outcome, there will likely not be the same bias inherent in the analyses for other outcomes. We describe this in greater detail below. Finally, we propose that in such outcome-wide studies various metrics that address issues of multiple testing be employed (Romano and Wolf, 2007; Mathur and VanderWeele, 2018a). While this suggestion is not new, we believe that if the outcome-wide template were embraced, the use of these various metrics in practice would become much more commonplace, and their effects on science, taken as a whole, much more substantial. In addition to the metrics, such as Bonferroni correction, that have been around for some time, we also introduce new metrics that are perhaps particularly well suited to outcome-wide studies in assessing the evidence, taken as a whole, for these various associations and potential effects. We also address some of the arguments against using such metrics and corrections.

In addition to addressing issues of bias from various sources, e.g. confounding or investigator discretion or multiple testing, we believe that our new template offers other additional and important advantages. It will allow for the expansion of knowledge much more rapidly over many more outcomes than does science carried out with the existing standard template. It will allow for the assessment of a single exposure on numerous outcomes simultaneously. We have argued elsewhere (VanderWeele, 2017a) that from a policy and public health perspective such an outcome-wide approach is important and that, ideally, we should be assessing the effects of exposures over numerous important outcomes, attempting as best as possible, to evaluate the effect of the exposure on human flourishing broadly construed. We should to the extent possible examine outcomes as diverse as happiness and life satisfaction, mental and physical health, meaning and purpose, character and virtue, close social relationships and financial security, amongst others (VanderWeele, 2017b). We believe this new template will help both in the advancement of knowledge and, we hope thereby, also the promotion of human flourishing. We will now describe this new outcome-wide longitudinal template in greater detail.

## 2. LONGITUDINAL DESIGNS FOR CAUSAL INFERENCE

### 2.1 Causal Inference Using Confounding Control

We will consider a setting in which we are interested in assessing the effect of some exposure or treatment A on a series of subsequent outcomes of interest $(Y_1,…,Y_K)$. With observational data, to draw causal inferences about the effect of exposure A on a particular outcome $Y_k$ certain assumptions need to be made about the comparability of the groups with and without exposure. Specifically, if a comparison of exposure groups is to be made and interpreted causally, it must be assumed that within strata of measured covariates C, the groups with and without exposure are comparable to one another in what would have occurred had each been in the alternative exposure group.

This assumption can be stated formally using counterfactual notation. We will begin our discussion of causal inference and confounder control with a single outcome $Y_k$ and will then discuss the implications of moving to a set of outcomes $(Y_1,…,Y_K)$. We will let $Y_k(a)$ denote the counterfactual outcome or potential outcome that would have been observed for an individual if the exposure A had, possibly contrary to fact, been set to level a. We say that the covariates C suffice to control for confounding if the counterfactuals $Y_k(a)$ are independent of A conditional on C, which we denote by notation $Y_k(a) \perp A \mid C$. The definition essentially states that within strata of C, the group that actually had exposure status A=a is representative of what would have occurred had the entire population with C=c been given exposure A=a. If this holds, we could use the observed data to reason about the effect of intervening to set A=a for the entire population.

This condition of no confounding for the effect of A on $Y_k$ conditional on C is sometimes, in other literatures, referred to using different terminology. It is sometimes in epidemiology also referred to as "exchangeability" (Greenland and Robins, 1986) or as "no unmeasured confounding" (Robins, 1992); in the statistics literature it is sometimes referred to as "weak ignorability" or "ignorable treatment assignment" (Rosenbaum and Rubin, 1983); in the social sciences it is sometimes referred to as "selection on observables" (Barnow et al., 1980; Imbens, 2004), or as "exogeneity" (Imbens, 2004). When this assumption holds and when we also have the technical consistency assumption that for those with A=a, we have that $Y_k(a)=Y$, then we can estimate causal effects (Pearl, 2009; VanderWeele, 2009), defined as a contrast of counterfactual outcomes, $E[Y_k(1) − Y_k(0)|c]$, using the observed data and associations. Specifically we then have that:

$$E[Y_k(1) − Y_k(0) \mid c] = E[Y_k \mid A = 1, c] − E[Y_k \mid A = 0, c]$$

The left hand side of the equation is the causal effect of the exposure on the outcome conditional on the covariates C=c. The right hand side of the equation consists of the observed associations between the exposure and the outcome in the actual observed data. If the effect of A on Y is unconfounded conditional on the measured covariates C, we can estimate causal effects from the observed data. The expression above is for causal effects on a difference scale, but if the effect of the exposure on the outcome is unconfounded conditional on covariates then one can likewise estimate the causal effect on the ratio scale from the observed data:

$$P[Y_k(1)=1|c] / P[Y_k(0)=1|c] = P[Y_k=1|A=1, c] / P[Y_k=1|A=0, c]$$

In general we will want to control for a sufficiently rich set of covariates C, related both the exposure and to the outcome, to make this assumption as plausible as possible. In the sections that follow we will discuss principles to guide the selection of these covariates C for a single outcome and then for a set of outcomes $(Y_1,…,Y_K)$.

**2.2 Longitudinal Data and Control for Baseline Outcome**

If the confounding control assumption is to be plausible it is first important that the actual data available be such that the exposure A temporally precedes the outcome. In most cases, then, cross-sectional data, in which all of the variables, A, $Y_k$, and C, are measured at the same time, will be useless for causal inference. For example, there is evidence that marital status is associated with higher levels of happiness; but with cross-sectional data it is impossible to know whether this is because marriage leads to happiness, or whether those who are happy are more likely to marry. In fact, there is evidence for both (Stutzer and Frey, 2006). The only way to begin to attempt to distinguish these possibilities is with longitudinal data, also sometimes called panel data, in which data is available for a group of the same individuals on multiple occasions. At least two waves of data will thus in general be a minimal requirement for attempting to draw causal inferences from observational data. Exceptions might occur when all of the data is collected at once but a particular exposure, and various covariates, are reported retrospectively. Such might be the case with, say, childhood experiences of parenting practices reported later in life. While from a data collection perspective this is cross-sectional, from the perspective of causal inference there is still a temporal ordering among the variables. One might still be worried about differential misreporting of childhood experiences affected by outcomes later in life, and we will turn to these considerations below, but from the perspective of temporality, the data would still have a longitudinal structure.

When such longitudinal data is available, it will often be important to control, whenever possible, for the outcome at or prior to the time of the baseline exposure assessment. For example, to attempt to evaluate the effect of marriage on subsequent happiness, it would be important to control for happiness levels earlier in life, prior to marriage, to attempt to rule out reverse causation – that the association between marriage and subsequent happiness is only due to happy people being more likely to marry. Such control for baseline outcome does not eliminate the possibility of reverse causation but helps to mitigate it (VanderWeele et al., 2016). Control for baseline outcome may not always be necessary if reverse causation can be ruled out on substantive grounds. For example, if one were attempting to assess the effect of parental religious service attendance when a child was age eight, on the child's subsequent voting behavior as young adult, it is unlikely that the eight-year-old child's sense of civic responsibility will have much effect on the parent's religious service attendance. However, in many cases control for baseline outcome will be important to make the confounding control assumption as plausible as possible; the baseline outcome may often be the strongest confounder affecting both the exposure and that same outcome subsequently. Thus, in addition to including a rich set of covariates related to the exposure and the outcome in the covariate set C, it will often be important to include in C also the baseline value of the outcome.

**2.3 Principles of Confounder Selection**

The question as to what variables to include the covariate set C can be a difficult one. Different disciplines often approach this question in different ways. Often in observational research in the biomedical sciences with large cohort dataset, an extensive set is included consisting of dozens of variables, sometimes including all of the data that are available. Sometimes in sociology and other social science disciplines it is more common to require justification for each and every covariate that is to be included. The goal for causal inference in any case is that, conditional on the final covariate set C, the groups with and without exposure are comparable.

Formal principles of confounder control have been articulated. It is well accepted that any common cause of the exposure and the outcome ought to be included in the covariate set C. It is also widely accepted that if we are interested in assessing the total effect of some exposure A on some outcome $Y_k$, then variables on the pathway from the exposure to the outcome ought not to be included as these might block some of the effect (Weinberg, 1993). Such a variable M on the pathway from the exposure to the outcome is a mediator of the effect, rather than a confounder (VanderWeele, 2015). Control for such a variable M might be appropriate if the goal were to assess the direct effect of the exposure on the outcome not through the mediator (Imai et al., 2010; VanderWeele, 2015), but if the goal is to assess the total effect of the exposure on the outcome then such variables ought not be controlled for. As an example, many analyses of the effect of education on happiness make adjustment for marital status, occupation and employment, and income. However, these variables are likely affected by and on the pathway from education to happiness; and indeed analyses using longitudinal data which take into account the fact that the temporal ordering of these variables, do find an effect of education on happiness, whereas analyses that control for these mediators do not (Cunado and de Garcia, 2012; Powdthavee et al., 2015). We should thus control for common causes of the exposure and outcome, but not for mediators between the exposure and outcome.

However, these basic principles are still consistent with several different practical approaches to thinking about what covariates to include. Pearl (2009) has derived a formal calculus for determining which set of covariates would suffice to control for confounding if knowledge were available of an entire causal diagram relating all variables to each other, including knowledge of all of the causal relationships among the covariates themselves. Such knowledge will often not be available. Various more practical proposals have been put forward. In statistics, it is sometimes recommended to control for all pre-exposure covariates (Rubin, 2008, 2009). While this may sometimes work well, it has been shown that there can be pre-exposure covariates the control for which increases, rather than decreases, bias, a phenomenon sometimes referred to as M-bias or collider-stratification bias (Sjolander, 2009; Ding and Miratrix, 2015). An alternative approach, sometimes articulated in epidemiology, is to control for all variables that are thought to be common causes of the exposure and the outcome (Glymour et al., 2008). While again this is intuitively appealing, there can be cases in which a particular measured covariate is not a common cause of the exposure and the outcome, but is instead e.g. on the pathway from an unmeasured common cause to the outcome, such that the measured covariate itself suffices to control for the confounding induced by the unmeasured common cause (VanderWeele and Shpitser, 2011). The principle of only controlling for common causes would thus not adequately control for confounding even though such control were possible using the measured covariates. The "pre-exposure" approach is in some sense too liberal with regard to the covariates that it includes, and the "common cause" approach is too conservative. An alternative is to attempt to include in the covariate set C any pre-exposure variable that is a cause of the exposure, or of the outcome, or of both. This has previously been referred to as the "disjunctive cause criterion" (VanderWeele and Shiptser, 2011). It can be shown that if this principle is used to determine what to control for in C, then if there exists any subset of the measured covariates that suffices to control for confounding then the subset selected by the disjunctive cause criterion will suffice as well (VanderWeele and Shiptser, 2011). This is not a property that is shared by the "pre-exposure" approach or the "common cause" approach. What is effectively discarded by the disjunctive cause criterion are those covariates that are neither causes of the exposure nor of the outcome. This disjunctive cause criterion is perhaps more similar to the approach sometimes employed in the social sciences of needing to justify the inclusion of each and every covariate as being a cause of either the exposure or the outcome. The difference here is arguably on which side is the burden of proof. With the disjunctive cause criterion, for the discarding of a covariate a case would need to be made that there is substantive and/or empirical evidence that the covariate in question is neither a cause of the exposure nor the outcome.

The disjunctive cause criterion has the attractive theoretical property noted above that if there exists any subset of the measured covariates that suffices to control for confounding then the subset selected by the disjunctive cause criteria will suffice as well. In practice, however, it may not perform as well if there is no subset of the measured covariates that would suffice. For example, when there is residual unmeasured confounding, it has been shown that control for an "instrumental variable" (e.g. a variable that is a cause of the exposure but is otherwise completely unrelated to the outcome except possible through the exposure) will often increase the bias already present due to unmeasured confounding (Pearl, 2010; Ding et al., 2017). It may thus be desirable in practice to exclude any known instrumental variables from the covariate set C. It has also been shown that control for a variable that is a proxy for an unmeasured common cause, will in many, though not all, contexts reduce bias and so it may be desirable to control for such variables as well (Ogburn and VanderWeele, 2013). A modified disjunctive cause criterion that might thus be more useful in practice could articulated as follows (VanderWeele, 2018): control for each covariate that is a cause of the exposure, or of the outcome, or of both; exclude from this set any variable known to be an instrumental variable; and include as a covariate any proxy for an unmeasured variable that is a common cause of both the exposure and the outcome.

**2.4 Common Confounders in Practice in Outcome-Wide Studies**

We have up until now focused on a relatively theoretical discussion of principles for confounder selection for assessing the effect of an exposure A on a single outcome $Y_k$. What are the implications for attempting to assess the effects of the exposure A on a broad range of outcomes $(Y_1,…,Y_K)$? If the goal were to select a single set of covariates C that sufficed to control for confounding for the effect of exposure A on each outcome $Y_k$ then one would want to include in C those covariates that were causes of either the exposure or of any outcome in $(Y_1,…,Y_K)$. This might well be a very broad set of covariates. It may in fact, bring one back to a set of covariates not very different from the pre-exposure approach. One only discards those variables that are thought to be a cause of *neither* the treatment nor of *any* outcome.

In principle one could apply the modified disjunctive cause criterion separately for each and every outcome $Y_k$ and select a different set of covariates for the assessing of each of the effects. We would argue against this approach on the following grounds: (1) As will also be discussed further below in Section 7.3 this can create temptation for investigators to fit, for each specific outcome, numerous different regressions controlling for different covariates and choosing the one they like best; this compromises the validity of the analysis. (2) There may be more disagreement over which covariates are a cause of a single outcome than which are a cause of any outcome; the former task may be considerably more difficult to correctly discern. (3) Often the outcomes will themselves affect one another; when this is the case a covariate which is principally the cause of one outcome may indirectly also be a cause of another outcome through the outcome for which it is a principal cause. (4) The analysis and reporting of results becomes more straightforward, as will be discussed below in Section 6.

If a broad range of outcomes are examined including, for example, those related to happiness and life satisfaction, mental and physical health, meaning and purpose, character and virtue, close social relationships, and financial outcomes (VanderWeele, 2017b), then the set of covariates selected for covariate control will also in general, ideally, be substantial. Any cause of any of these outcomes, measured prior to exposure, should be included. This would thus also ideally include, as per the discussion in Section 2.3 above, baseline values of all outcomes whenever appropriate. Doing so will of course necessitate rather large sample sizes in practice and we would thus encourage these outcome-wide analyses principally for large cohort datasets.

Often different disciplines place greater or less emphasis on particular sets of specific covariates. It can be instructive to consider the whole range of these when carrying out outcome-wide analyses. In most disciplines control is made whenever possible for various demographic characteristics such as race, gender, age, and marital status. In biomedical research, effort is also made to additionally control for various measures of physical and mental health as well as for health behaviors; at the very least effort is made to control for exercise, smoking, alcohol consumption, self-rated physical health or either various health conditions or their number, and depression. We would argue that these variables ought to be included, whenever possible, in outcome-wide analyses. Health goes on to affect many other outcomes also. Within economics, effort is often made to additionally control for measures of income, education, and employment. These too should be included, when possible, for covariate control in outcome-wide studies. Within sociology effort is often made to additionally control for social integration and support, quality of neighborhood, and religious practice; within political science, political affiliation is often associated with numerous outcomes. Much of the more prominent research in psychology is experimental rather than observational, but within psychology there is strong evidence of the following variables affecting numerous outcomes: life-satisfaction/happiness, loneliness, parental warmth, purpose or worthwhile activities, and the "big 5 personality" traits (Gosling et al., 2003). We believe all these too should be controlled for, whenever possible, in outcome-wide analyses. A list of these covariates is summarized in Table 1. Of these, we believe that those that are perhaps most frequently neglected in observational research intended to assess causal effects are (i) parental warmth during childhood which has been shown to affect numerous outcomes; (ii) the "big 5" personality traits (extraversion, conscientiousness, agreeableness, neuroticism, openness) as these likewise affect numerous outcomes; (iii) political affiliation; and (iv) religious service attendance which is likewise strongly associated with a very broad range of outcomes (Koenig et al., 2012; VanderWeele, 2017c). The first three of these are perhaps less often available in large cohort datasets, but could be, and we believe should be, more often in the future included in the form of simple measures (e.g. Gosling et al., 2003). The fourth of these, religious service attendance, often is available, and could, and we believe should, be controlled for as a covariate more frequently. It is most often a stronger predictor than other affiliation or private practice religious/spiritual variables (Koenig et al., 2012; VanderWeele, 2017c). Perhaps more controversially, measures of intelligence have been shown to be associated with a number of outcomes (Nisbett et al., 2012); however, such data are currently rarely available in most cohort studies.

The list given here is not meant to be exhaustive but only indicative of what are major causes of many outcomes across these various disciplines are and therefore helps inform what covariates one might aim to adjust for in confounding control in an outcome-wide analysis. The list can be daunting. Very few datasets will have information on all of these. Thankfully, as discussed further below in Section 3, residual unmeasured confounding that is generated by an unmeasured variable will only create bias to the extent that it is orthogonal to all measured covariates (VanderWeele, et al., 2018). Often when the set of measured covariates is rich, the residual confounding generated by an unmeasured covariate will be small. It can be instructive to go through the measured covariates and omit them one at a time. If the set of measured covariates is rich then even the omission of what are otherwise important and highly predictive variables, such as race or income, will not change effect estimates all that much when omitted, since the *residual* confounding, *conditional on* all of the other measured covariates, ends up being quite small; the other measured covariates control for most of it. We will return to this point below in our discussion of unmeasured confounding in Section 3. Nevertheless, because one can never be certain that the measured covariates suffice to control for confounding or that the residual unmeasured confounder is small, it is important to assess the robustness of one's conclusion and effect estimates to potential unmeasured confounding, and therefore sensitivity analysis for unmeasured confounding and other biases will be important. This is the topic of Section 3 below and we strongly encourage the use of the robustness metrics there in all outcome-wide studies.

**2.5 Timing of Confounders**

Another consideration that should be taken into account when making decisions about confounder selection based on substantive knowledge is that of covariate timing. It was noted above that for estimation of total effects, rather than direct effects, we do not want to make adjustment for variables that may be on the pathway from the exposure to the outcome. To avoid this, we often refrain from adjusting for covariates that occur temporally subsequent to the exposure. In many two-wave longitudinal studies, the exposure and covariates are all assessed at one time and the outcome is assessed at a subsequent time. However, in a number of cohort studies, data is collected on all exposures, covariates, and outcomes repeatedly across each wave, perhaps once per year, or once every two years, for many years or even decades. Such designs can help make more informed confounder selection decisions based on the temporal ordering of the data. One difficulty with studies in which the exposure and potential confounding covariates are all assessed at the same time is that it can be difficult to determine whether a covariate assessed at the same time as the exposure may in fact be affected by it and thus be a mediator rather than a confounder.

Consider, for example, a study intended to assess the effect of physical activity on cardiovascular disease. Body mass index (BMI) might be available as a covariate and it may then thought to be important to control for BMI as a confounder. However, it is of course also conceivable that BMI is on the pathway from physical activity to cardiovascular disease and that control for it may block some of the effect of physical activity. Conversely, it may also be the case that BMI itself affects both subsequent physical activity and subsequent incidence of cardiovascular disease. Someone with a very high BMI may have more difficulty regularly exercising. Thus it is possible that BMI is both a confounder (for the effect of subsequent physical activity) and also a mediator on the pathway from prior physical activity to cardiovascular disease. It is thus difficult to know whether or not to adjust for BMI if both BMI and physical activity are measured at the same time. We cannot adequately distinguish in this setting between confounding and mediation (VanderWeele, 2015). If, however, BMI is available repeatedly over time then it may be possible to control for BMI in the wave of data that is prior to the wave that uses exercise as the primary exposure. This would better rule out the possibility that the BMI variable used in the analysis is a mediator; if its measurement precedes that of physical activity by a year then it is more reasonable to interpret it as a confounder. When multiple waves of data are available it may thus be desirable to control for the covariates in the wave prior to the primary exposure of interest.

This will not always a reasonable option for potentially two reasons: either because there are only two waves of data are available (one for the exposure and covariates, and one for the outcome), or alternatively because, although a wave of prior covariate data is available, it may be temporally too far prior to the exposure measurement to be of adequate use for confounding control. For example, if the prior wave of data is ten years prior to the exposure measurement, it will be much less effective at ruling out confounding than if it were one year prior. Depending on how far back the prior wave of data is, there will be a trade-off between the potential for residual unmeasured confounding, if the wave is too far back, versus the danger of controlling for a variable that is a mediator, if the covariates for which control is made are contemporaneous with the exposure.

It is also of course possible to carry out sensitivity analysis of the timing of confounder measurement, and to compare the results when confounders are controlled for contemporaneously with the exposures versus when they are controlled for in the prior wave (e.g. Danaei et al., 2013; Garcia-Aymerich et al., 2014). When contemporaneous control for the covariates is made, the danger of adjusting for mediators, especially when numerous covariates are included in the model as suggested above in section 2.4, can be substantial. It may thus also be desirable as an additional sensitivity analysis to go through each of the covariates and consider, substantively, whether each covariate is more likely to immediately affect the exposure, or whether the covariate is more likely to immediately be affected by the exposure, and, in a supplementary analysis only control for the former set of covariates. Ideally, however, designs would allow for covariate control shortly prior to the exposure measurement.

**2.6 Control for Prior Exposure**

A final issue concerning covariate control concerns potentially controlling also for prior values of the exposure variable itself. This only makes sense when the exposure varies over time. For an exposure such as exercise, or employment, or religious service attendance, the exposure itself may change across the waves of data. In such settings one can attempt to assess the effects of an exposure trajectory on final outcomes. The confounding control assumptions required to assess the effects of time-varying exposures are more complex in this setting and are described elsewhere (Robins, 1992, Robins et al., 2000; Robins and Hernán, 2009; Hernán and Robins, 2018); we will also comment on this setting further below in Section 8.4. Here we will continue to focus on the setting of assessing the effect of an exposure at a single point in time.

In this setting, if the exposure can itself change over time then it may be desirable to control also for the value of the exposure in the prior wave of data. This can be desirable for a number of reasons. First, it facilitates the interpretation of the effect estimate as a change in the exposure from e.g. absent to present. Without control for prior exposure, such an interpretation is justified only if the prior value of the exposure is independent of the outcome conditional on the baseline exposure and measured covariates. Control for prior exposure might be done either by including it as a covariate or by stratifying the analysis by prior exposure status. Second, control for prior exposure can help further rule out reverse causation: if the value of the outcome two periods prior to the exposure affects both the baseline exposure independently of the outcome one period prior, and further affects the final outcome independently of the exposure and the outcome one period prior, then simple control for the baseline outcome as suggested in Section 2.2 will not suffice to rule out reverse causation, whereas control also for baseline exposure can, in many settings, further rule out reverse causation (VanderWeele et al., 2016). Third, control for prior exposure can also help further rule out other forms of unmeasured confounding. This is so because, if control is made for prior exposure then, for an unmeasured confounder U to explain away an observed exposure-outcome association, the unmeasured confounder would have to be associated with both the outcome and the baseline exposure, independent of the prior level of exposure. Thus, in Figure 1, both of the dashed arrows to the baseline exposure and to the final outcome would have to be present and substantial to induce considerable confounding bias. Consider, for example, a study examining the effects of religious service attendance on depression; suppose no control was made for the "big five" personality traits. It is known that conscientiousness is associated with both higher religious service attendance and lower depression; if this were not adjusted for, it might be thought to induce confounding. However, if control is made for prior level of the exposure then for conscientiousness to explain away the exposure-outcome association, conscientiousness would have to be substantially associated with the religious service attendance exposure, independent of prior religious service attendance, and this may be less plausible. Fourth and finally, control for prior exposure can help rule out instances in which initiation of the exposure itself may, in the short-run, have harmful consequences and thereafter look beneficial (Danaei et al., 2012; Hernán, 2015). In the epidemiologic literature, controlling for prior exposure is sometimes referred as an analysis assessing the effects of "incident exposure" rather than "prevalent exposure" (Danaei et al., 2012; Hernán, 2015) and, for the reasons above, this can sometimes be preferable.

However, such control for past exposure may not always be needed or preferable. Certain exposures may be relatively stable over time; for example, while parenting practices for a given parent can change over time, for most, they may be relatively stable and using a single exposure assessment may be sufficient. For other exposures, such as an introduction of a job training program that is new to a community, it is possible that no one has previously been exposed, and thus that there is no data on prior exposure, but also no need to adjust for it, since its values is effectively zero for all study participants in all prior waves. For exposures that are relatively stable there may be very little or almost no change across waves in which case the baseline exposure and the past exposure will be almost entirely collinear. If changes do occur but are rare then very substantial sample sizes may be needed to be able to control for past exposure (for example, in an analysis to assess the effects of religious service attendance on mortality, Li et al. (2016) found only slight changes in religious service attendance categories across four years; however, with a sample size of over 74,000, it was still possible to fit models that controlled for past service attendance). In other cases, it might also be undesirable to control for past exposure when the prior wave of data for which the exposure is available was in the distant past, as this can potentially introduce the types of biases that arise with time-dependent confounding for time-varying exposures (Robins, 1992, Robins et al., 2000; Robins and Hernán, 2009; Hernán and Robins, 2018). It may be more reasonable to control for prior exposure when it is a year or two prior to baseline exposure than when it is ten years prior and thus likely altered considerably in the intervening ten years as well.

However, when prior exposure data is available in the relatively recent past and when the exposure itself changes with sufficient frequency and sample sizes are such as to allow for prior exposure as an additional covariate, it can be desirable, for the reasons mentioned in the previous paragraph, to add it as a covariate as well.

A hierarchy of evidence from different study designs for assessing causality might thus be formulated (VanderWeele et al., 2016). First, at the weakest level of the hierarchy are cross-sectional designs and analyses; these will in general contribute little evidence for causality unless a clear argument can be made for the temporal ordering of the exposure preceding the outcome and control can be made for confounding variables that likewise temporally precede the exposure and outcome. Second, longitudinal designs in which the exposure clearly precedes the outcome and in which control can be made for a rich set of baseline covariates that potentially confound the relationship between the exposure and the outcome have more potential to contribute some evidence for causality. Third, if control can also be made for prior measures of the outcome this strengthens the evidence further as control for prior or baseline outcome can help rule out reverse causation. Fourth, if control can also be made for prior exposure this strengthens the evidence yet further for the reasons given above. Finally, a randomized trial of the exposure generally provides, at least in the absence of complications such as non-compliance and drop-out, the strongest evidence for a causal relationship. In most, though not all cases, we believe that at least level three of the hierarchy above (control for baseline outcome) needs to be achieved to have the potential to contribute substantially to evidence for causality, unless a compelling case can be made for ruling out reverse causation on substantive grounds. Evidence for a causal relationship depends of course also on other details of the design, the size of the study, the magnitude of the effect estimate, the richness of the covariate data, the quality of measurements, and various other factors, all which all must be carefully evaluated, and which are discussed further in Section 3 below. Nevertheless, questions of temporality in study design and controlling for prior values of outcome and possibly exposure ought to be given considerable weight in assessing evidence for causality.

**2.7 Outcome-Wide Regression Models and Estimation**

The discussion in Section 2.2-2.6 above was all oriented around study design considerations and choice of covariate control. Once these are in place the proposed statistical analysis for an outcome-wide study is relatively straightforward. One could, for example, for each continuous outcome, $Y_k$, fit a linear regression model of $Y_k$ on exposure A and the covariates C that were selected using the principles discussed above (including in C, when applicable, prior values of outcome and exposure):

$$E[Y_k|a,c] = \alpha_k + \beta_k a + \gamma_k'c$$

and, for each dichotomous outcome, fit the analogous logistic regression:

$$\text{logit}[P(Y_k=1|a,c)] = \alpha_k + \beta_k a + \gamma_k'c$$

and, for each count outcome, to fit the analogous Poisson regression:

$$\log(E[Y_k|a,c]) = \alpha_k + \beta_k a + \gamma_k'c$$

and likewise for other regression models that may be of interest. For each outcome k, provided the confounding control assumption holds that $Y_k(a) \perp A | C$, the coefficient $\beta_k$ in each model will provide a consistent estimate of the causal effect of exposure A on outcome $Y_k$ on the relevant scale corresponding to the regression model being used. For example, for a linear regression model $E[Y_k|a,c] = \alpha_k + \beta_k a + \gamma'c$, we have that, provided $Y_k(a) \perp A | C$, then $\beta_k = E[Y_k(1) - Y_k(0) | c]$. For a rare outcome such that odds ratios approximate risk ratios, the causal risk ratio can be obtained by exponentiating the coefficient in the logistic regression model so that $\exp(\beta_k) = P[Y_k(1) = 1|c] / P[Y_k(0) = 1|c]$. It may be desired to restrict the use of logistic regression models to dichotomous outcomes that are relatively rare so that the odds ratio approximate the risk ratio; otherwise such odds ratios can vastly exaggerate the corresponding risk ratio (cf. VanderWeele, 2017d). With common outcomes, other estimation strategies to obtain risk ratios directly, such as a modified Poisson regression or a log-binomial model, might be used (Yelland et al., 2011; Knol et al., 2012).

While global inference on regression coefficients for different outcomes could alternatively be conducted using multivariate regression (Johnson and Wichern, 2002) or with a "seemingly unrelated regressions" generalization (Zellner, 1962), these approaches at best only modestly improve efficiency compared to that achieved in K separate linear regression models; when the design matrix is shared across models, as we suggested be done above in Section 2.4, coefficient estimates are identical to those using ordinary least square estimation (Oliveira and Teixeira-Pinto, 2015). Conducting K separate regression models will thus often suffice for these outcome-wide analyses.

An alternative analytic approach would be to carry out propensity score analyses (Rosenbaum and Rubin, 1983b) for each outcome either via matching or subclassification (Rosenbaum, 2002). Because propensity score subclasses and matches are formed without reference to the outcome, the same subclassification or matched sets can be used for all outcomes, thereby also more easily facilitating automation of the analyses when a large number K of outcomes are being examined. Other matching approaches, not based on propensity scores, could also be used. However matching approaches in which decisions about covariate matching are based on the outcome (e.g. Iacus et al., 2012), while useful in the contexts of a single outcome, may be more difficult to apply outcome-wide as the decisions would have to be made separately for each outcome.

Alternatively, doubly robust estimators or machine learning or high dimensional covariate selection algorithms (van der Laan and Rose, 2011, 2018; Belloni et al, 2014; Schuler and Rose, 2017) could be used to obtain effect estimates. We believe these approaches are potentially promising in the outcome-wide setting as well, but further work on determining when sample sizes are adequate for the desirable asymptotic properties of these estimators to apply is needed. Other approaches are also available for inference when translating effects on multiple outcomes to a common scale, using mean-variance and median-interquartile range based standardizations (Kennedy et al, 2018). The focus of this paper is on the outcome-wide longitudinal design itself and the approach is compatible with a number of different statistical modeling options.

## 3. E-VALUES FOR UNMEAURED CONFOUNDING AND OTHER BIASES

In the previous section we considered causal inference for outcome-wide studies using confounding control. The assumption that the measured covariates C suffice to control for confounding is a strong one and will, even at best, only hold approximately. It is thus important to assess the robustness of causal effect estimates to violations of this assumption. Sensitivity analysis techniques for unmeasured confounding are useful in this regard. A variety of techniques are available (e.g. Rosenbaum and Rubin, 1983a; Rothman et al., 2008; Lash et al., 2009). Here we will consider a relatively simple approach that we believe is particularly well-suited to outcome-wide studies, and consists of reporting a metric of robustness to unmeasured confounding called the E-value (VanderWeele and Ding, 2017). In Section 3.1 we will discuss this E-value approach to assessing unmeasured confounding; in section 3.2 we will discuss the implications of such sensitivity analysis for assessing evidence for causality; and in section 3.3 we will discuss other forms of bias, beyond unmeasured confounding, that may threaten outcome-wide analyses.

**3.1 Sensitivity Analysis for Unmeasured Confounding**

The E-value is a metric that can be used to assess robustness of longitudinal associations to potential for unmeasured confounding. As such it is a measure relevant to assessing *evidence* for causality in observational research. The E-value metric itself arises from sensitivity analysis for unmeasured confounding. The formal derivation of the E-value relies on two parameters (Ding and VanderWeele, 2016). We will begin our development with a binary outcome $Y_k$ and then comment upon other types of outcomes as well. The observed exposure-outcome association on the risk ratio scale, conditional on covariates C, is given by

$$RR_{obs} = \frac{P(Y_k = 1|A = 1, c)}{P(Y_k = 1|A = 0, c)}$$

The association, conditional on C, but adjusted also for some set of unmeasured confounders U would be:

$$RR_{true} = \frac{\sum_u P(Y_k = 1|A = 1, c, u)P(u|c)}{\sum_u P(Y_k = 1|A = 0, c, u)P(u|c)}$$

If covariates (C,U) suffice to control for confounding of the effect of A on $Y_k$, then the latter expression $RR_{true}$ can be interpreted as the causal risk ratio of A on $Y_k$ conditional on C i.e. $P[Y_k(1) =1|c] / P[Y_k(0) =1|c]$. Consider now the following two sensitivity analysis parameters (Ding and VanderWeele, 2016; VanderWeele and Ding, 2017):

$$RR_{UY_k} = max\left\{\frac{max_u P(Y_k = 1|A = 1, c, u)}{min_u P(Y_k = 1|A = 1, c, u)}, \frac{max_u P(Y_k = 1|A = 0, c, u)}{min_u P(Y_k = 1|A = 0, c, u)}\right\}$$

$$RR_{AU} = max_u \frac{P(U = u|A = 1, c)}{P(U = u|A = 0, c)}$$

Essentially, $RR_{UY_k}$ is the maximum effect that U can have on $Y_k$, conditional on C=c, comparing any two categories of U, for either the exposed or unexposed; and $RR_{AU}$ is the maximum risk ratio relating the exposure to any particular level of U, conditional on C=c. Ding and VanderWeele (2016) derived the following sharp bound:

$$\frac{RR_{obs}}{RR_{true}} \leq \frac{RR_{UY_k} \times RR_{AU}}{RR_{Y_k} + RR_{AU} - 1}$$

so that $\frac{RR_{UY_k} \times RR_{AU}}{RR_{UY_k} + RR_{AU} - 1}$ was the maximum bias (comparing the ratio of the observed association adjusted for C, to the true association adjusted also for U) that could be generated by such an unmeasured confounder. It was then further derived that for the unmeasured confounder(s) to shift the observed risk ratio to the null of 1, if one wanted both RR$_{UYk}$ and RR$_{AU}$ to be as small as possible, then the minimum they could both be (which was what was called the E-value) was:

$$E-value = RR_{obs} + \sqrt{RR_{obs}(RR_{obs} - 1)}$$

For risk ratios that are protective rather than causative, the inverse of the observed risk $RR_{obs}$ is taken before applying the E-value formula above.

The E-value is thus straightforward to calculate from the observed risk ratio. As an example, the E-value for an observed risk ratio of RR=1.3 is 1.92. Thus with an observed risk ratio of 1.3, an unmeasured confounder that was associated with both the exposure and the outcome by risk ratios of 1.92-fold each, conditional on the measured covariates, would suffice but weaker confounding would not (where the strength of confounding is defined by the bias factor $\frac{RR_{UY_k} \times RR_{AU}}{RR_{UY_k} + RR_{AU} - 1}$). As other examples, the E-value for a risk ratio of RR=1.1 is 1.43; the E-value for a risk ratio of RR=1.5 is 2.36; the E-value for a risk ratio of RR=2 is 3.41. As can be seen from the formula above, the E-value will always be larger than the observed risk ratio. The relationship is highly non-linear for modest values of the risk ratio that are slightly above 1.

An E-value for the confidence interval can also be report to determine the minimum confounding that would be needed to shift the confidence interval to include the null. The E-value for the confidence interval is obtained by assigning the E-value of 1 if the confidence interval contains the null and otherwise applying the E-value formula to the limit of the confidence interval that is closest to the null. The E-value for the confidence interval has the interpretation that "Across repeated samples, at least 95% of the time it is the case that: if the actual confounding parameters RR$_{UYk}$ and RR$_{AU}$ are both less than the E-value for the confidence interval that was calculated, then there will be a true effect in the same direction as the observed association." (VanderWeele et al., 2018). In outcome-wide analyses, for each outcome, we recommend reporting the E-value both for the estimate and for the limit of the confidence interval closest to the null. This simple metric gives the investigator and reader a sense as to how robust, or sensitive, effect estimates are to unmeasured confounding and this robustness or sensitivity can be seen to vary across outcomes.

Several points are important in the interpretation of the E-value. First, the confounding associations RR$_{UYk}$ and RR$_{AU}$ are both conditional on the measured covariates C so that the confounding associations RR$_{UYk}$ and RR$_{AU}$ reflect residual confounding not captured by the measured covariates C. It is the association between U and both Y$_k$ and A, *independent of C*, that is relevant here. Second, the inequality holds for any U and thus the results are relevant for any set of covariates U such that the effect of A on Y$_k$ is unconfounded conditional on (C,U). One could thus define the parameters RR$_{UYk}$ and RR$_{AU}$ for each possible U such that (C,U) suffice to control for confounding and then take the minimum over U of the resulting bias. The E-value calculated as $RR_{obs} + \sqrt{RR_{obs}(RR_{obs} - 1)}$ is then the minimum strength of association on the risk ratio scale that *any and every* unmeasured confounder, that suffices along with C to control for confounding, would have to have with both the exposure and the outcome, above and beyond the measured covariates, to explain away the observed exposure-outcome association. Approximate versions of such statements are also possible if confounding control is not exact (VanderWeele et al., 2018).

The bias analysis and E-value calculations above are in fact applicable to the setting of multiple unmeasured confounders. The confounding parameters $RR_{UY_k}$ and $RR_{AU}$ are then simply interpreted respectively as the maximum effect that U can have on $Y_k$, conditional on C, comparing any two categories of the entire vector of unmeasured confounders U, for either the exposed or unexposed; and $RR_{AU}$, is the maximum risk ratio relating the exposure to any particular level of the entire vector U, conditional on C. In such settings, large values of $RR_{UY_k}$ and $RR_{AU}$ may not be particularly implausible. While an E-value of 5 say, may seem, when considering a single confounder, to require very substantial confounding associations and it is perhaps unlikely a single unmeasured confounder could increase the probability of the outcome by 5-fold, above and beyond the measured covariates, an increase of that magnitude may not be as implausible if one is considering a whole group of potential unmeasured confounders. The effect comparing the most favorable values of a set of confounders U to the least favorable values of that set U might plausibly increase the probability of the outcome by 5-fold, perhaps even above and beyond the measured covariates. However if this is indeed so, one should perhaps question whether the data available are in fact adequate to get a reasonable estimate of the causal effect at all. If it is known in advance that there are not just one, but numerous known unmeasured confounders, strongly associated with the outcome(s) and exposure and independent of the measured covariates, then arguably this is not a good study setting in which to attempt to draw conclusions. It may then be best to pursue other more adequate data sources.

The initial estimate for the causal effect should of course be adjusted for as many measured confounders as possible according to the principles of Section 2 above. A large E-value is only strong evidence for a true causal effect if the set of measured covariates adjusted for plausibly controls for much of the confounding. Said another way, the design of the study, and the collection of data on measured and known confounders, is essential in whether a large E-value gives strong evidence or not.

The E-value is in fact a conservative measure of robustness to unmeasured confounding insofar as, if the parameters $RR_{UY_k}$ and $RR_{AU}$ are in fact as large as the E-value, then it is possible to construct scenarios in which an unmeasured confounder U with those parameters would suffice to bring the observed association down to the null. However, there are also many other scenarios in which the actual unmeasured confounder has confounding parameters $RR_{UY_k}$ and $RR_{AU}$ that are equal to the E-value and yet the unmeasured confounder would not suffice to reduce the observed association to the null. The inequality for the maximum bias $\frac{RR_{obs}}{RR_{true}} \leq \frac{RR_{UY_k} \times RR_{AU}}{RR_{UY_k} + RR_{AU} - 1}$ is an inequality, not an equality. The inequality is sharp in that it is always possible to construct a variable U with those confounding associations that attains the bound, but, with an actual unmeasured confounder, the bias will often be less. This is especially the case when, for example, the unmeasured confounder is rare (Ding and VanderWeele, 2016). The E-value essentially assumes that the distribution of U is as unfavorable as possible.

The development above applies for a binary outcome using risk ratios. However, using various approximate conversions often employed in the meta-analysis literature between odds ratios and standardized effect sizes for continuous outcomes (Hasselblad and Hedges, 1995; Borenstein et al., 2009), and between odds ratios and risk ratio (VanderWeele, 2017d), one can obtain approximate E-values for other outcome scales.

For a continuous outcome, with a standardized effect size "d" (obtained by dividing the mean difference on the outcome variable between exposure groups by the standard deviation of the outcome) and a standard error for this effect size $s_d$, an approximate E-value can be obtained (VanderWeele and Ding, 2017) by applying the approximation RR≈exp(0.91×d) and then using the E-value formula above $E-value = RR_{obs} + \sqrt{RR_{obs}(RR_{obs} - 1)}$. An approximate confidence interval can be found using the approximation, (exp{0.91×d – 1.78×$s_d$}, exp{0.91×d + 1.78×$s_d$}) and then obtaining the E-value for the confidence interval. Approximate E-values for other effect measures such as odds ratios, hazard ratios, and risk differences can also be obtained (see VanderWeele and Ding, 2017). An online E-value calculator (https://evalue.hmdc.harvard.edu/app/) and R-package (Mathur et al., 2018) are also available to obtain these E-values automatically. With E-values for these other effect scales, the approach relies on additional assumptions and approximations (unlike for risk ratios). Other sensitivity analysis techniques have been developed for continuous outcomes (e.g. Lin et al., 1998; Imbens, 2003; VanderWeele and Arah, 2011), but these likewise require additional assumptions. An advantage of the E-value approach is that it provides a common, at least approximate, scale for assessing robustness to unmeasured confounding across different types of outcomes, though the E-value itself must always be interpreted within the context of the particular exposure, outcome and set of covariates under consideration (VanderWeele and Ding, 2017).

**3.2 Skepticism with Regard to Causal Effects from Selection on Observables**

In certain circles and within economics especially, there can be considerable skepticism that it is ever possible to provide substantial evidence for causation using regression models with the type of "confounding control" or "selection on observables" assumptions that were discussed in Section 2. While we believe that a critical approach needs to be taken to the interpretation of such regression analyses, we also believe that such extreme skepticism, when applied universally, is misguided. We believe that the difference in levels of skepticism about the plausibility of the selection on observables assumption across disciplines arises in part because of the nature of the data often available and also in part because of the different contexts of the systems and phenomena under study. However, we also believe that the approach we are advocating as laid out in Section 2 and Section 3.1 has the potential to provide substantial evidence, contrary to the extreme skepticism sometimes expressed especially in economics.

With regard to the issue of data availability, many observational studies of secondary data in economics have relatively little covariate information; the dataset being used may have been collected for one purpose but is being used for another. If the initial set of measured covariates that is available for control is weak or limited, then skepticism is certainly warranted. In contrast, however, in many biomedical studies, much richer covariate data is available. Often large cohort studies to examine the determinants of health are designed specifically with that goal in mind, with careful thought being given to what variables might confound the relationships between the exposures and health outcomes under study. Often the covariate data is very rich indeed. Of the covariates discussed in Section 2.4, in many large biomedical cohort studies, data is available on almost of all these with the exception of the "big five" personality traits and political affiliation. Some of the differences in levels of skepticism may thus be due to the availability of covariate data and, thus also, the plausibility of the "selection on observables" assumption. However, some of the difference in levels of skepticism may also have to deal with the different nature of the phenomena being studied across disciplines. In many economic contexts it is assumed that agents have some degree of information about their own potential outcomes that is not available in the data for which measurements are available, and that the agents use this information to select into the treatment or exposure groups. For example, decisions about occupation may be made based on an agent's own assessment as to where they are likely to be successful. In contrast, in a number of biomedical settings, the patient or participant may not have analogous information; it may be that the patient's physician is the principal decision-maker concerning which treatment may be best, and that the information available to the physician is in fact roughly the same information available in the data to the researcher. Hence, some of the discrepancy in the degree of skepticism about causal inference through covariate adjustment may arise from the different objects of study. Different levels of skepticism may be merited by different disciplines.

However, in addressing the extreme skepticism with regard to causal inference using covariate adjustment, several further points merit attention. First, as noted above, in some contexts at least relatively rich covariate data may be available. Second, when rich covariate data is available, then, even if there are seemingly important unmeasured confounders, the measured covariates may in fact adjust for a substantial portion of the unmeasured confounding leaving relatively little residual confounding remaining. It was noted above that an unmeasured variable will only introduce residual unmeasured confounding to the extent that it is associated with both the exposure and the outcome, independent of all of the measured covariates. It was thus also noted that if the set of measured covariates is rich then even the omission of what are otherwise important and highly predictive variables, such as race or income, will often not change effect estimates all that much when omitted, because the *residual* confounding is *conditional on* all of the other measured covariates. Third, using the E-value metric or other sensitivity analyses techniques, it may sometimes be established that very substantial residual unmeasured confounding would be needed to explain away a covariate-adjusted exposure-outcome association. A well designed longitudinal study with control for a rich set of covariates, along with control for prior outcome and exposure, that is accompanied by a large E-value, may constitute very strong evidence indeed for a causal effect of the exposure on an outcome.

**3.3 Sensitivity Analysis for other Types of Bias**

Of course unmeasured confounding does not represent the only threat to the validity of analyses assessing causal effects. Biases can arise from measurement error; biases can arise from missing data; biases can arise censoring or selection on or restriction to the study sample based on a variable affected by the exposure or outcome. These biases too can be very important. We will briefly discuss these various biases, specifically as they relate to outcome-wide analyses. We believe, for the reasons given below, that robustness to unmeasured confounding, using the E-value, or some other metric, should always be carried out in outcome-wide analyses, but that these other forms of bias may, or may not, merit further attention depending on the context.

Measurement error can be a threat to analyses intended to assess causal effects. Non-differential measurement error, in which the measurement error of the exposure (or outcome) does not depend on the outcome (or exposure, respectively) will often, though not always, result in estimates that are biased towards the null (Bross, 1954; Weinberg et al., 1994; VanderWeele and Hernán, 2011). If the non-differential measurement error is in the exposure, it may be relatively straightforward to apply measurement correction approaches outcome-wide (Carroll et al., 2006; Rothman et al., 2008; Lash et al., 2009). If the measurement error is in the outcome(s), and some of those outcomes are binary, then applying correction approaches outcome-wide will be more challenging as each outcome will require distinct correction parameters (Carroll et al., 2006; Rothman et al., 2008; Lash et al., 2009). However, even if non-differential measurement error is ignored, each of the effect estimates will thus often constitute conservative estimates, at least with respect to measurement error. Differential measurement error, in which the measurement error in the exposure depends on the outcome, or the measurement error in the outcome depends upon the exposure, may be more of a threat to outcome-wide analyses. Such measurement error will often bias effect estimates away from the null. Analogous metrics to the E-value but for measurement error are available (VanderWeele and Li, 2018). However, in most cases these effectively amount to requiring that for differential measurement error to explain away the association the effect of the outcome on the exposure measurement independent of the true exposure (or the effect of the exposure on the outcome measurement independent of the true outcome) must be at least as large as the effect estimate (VanderWeele and Li, 2018). The effects estimates and confidence intervals themselves in an outcome-wide study thus constitute the relevant bounds concerning the minimal differential measurement error needed to explain away the association and so no further reporting is needed.

In cases in which the restriction of the sample is made based on a variable affected by the exposure or outcome, the biases that are induced can be substantial indeed. Metrics analogous to the E-value are likewise available for this setting as well (Smith and VanderWeele, 2018). However, for such selection bias, unlike for the E-value for unmeasured confounding, in many cases the magnitude of the associations of the bias parameters required to explain away the observed exposure-outcome association will in fact be smaller, rather than larger than, the observed exposure-outcome relationship itself. We would thus caution against outcome-wide analyses when selection bias is thought to be substantial. Depending on the nature and type of selection bias, more careful and thoughtful assessment of each outcome may be needed.

Fortunately, in contrast to unmeasured confounding, differential measurement error and selection bias due to restriction will not be major threats in all observational studies. While measurement error may be pervasive, differential measurement error will be more rare. Selection bias due to restriction may be present in some studies, but in many, it is not a substantial concern. In contrast, however, whenever observational data are used to draw causal inferences, unmeasured confounding will be a concern. We thus recommend always reporting the E-value for unmeasured confounding (or using some other sensitivity analysis) in all outcome-wide studies, and then dealing with measurement error and/or selection bias due to restriction, when necessary, along the lines of the principles suggested above.

We will conclude this section with some discussion of missing data. In a number of large cohort datasets, data is missing on certain covariates for some individuals, other covariates for other individuals, the exposure for some, and the outcome for others, without any clear patterns with regard consistent missingness. In such settings, we believe that multiple imputation (Little and Rubin, 2014) can be an effective way to address such missing data issues. However, given that the proposed outcome-wide analyses are intended to examine numerous outcomes at once and it is therefore not possible to give the same degree of attention to any single exposure-outcome analysis, we would advise caution with using the outcome-wide approach when missing data is extensive (e.g. considerably more than 10% for any given covariate or exposure or outcome). We would also recommend comparing estimates obtained by multiple imputation with a complete case analysis. Similarity in results may provide reassurance (though does not guarantee) that the missing data itself is not causing substantial bias. Major discrepancies between the complete cases analyses and the multiple imputation results may indicate that the missing data is indeed a threat to the effect estimates and that further sensitivity analyses for missing data, including those that consider missing-not-at-random scenarios, may be desirable. In such cases, it may be better to abandon the outcome-wide approach and consider each outcome individually while more carefully addressing issues of missing data.

In summary, we believe that in outcome-wide analyses robustness or sensitivity to unmeasured confounding can be addressed in a relatively straightforward way, outcome-wide, using the E-value. Outcome-wide analyses subject to non-differential measurement error will often yield conservative results; when correction for non-differential measurement error is desired, it will be more feasible to carry this out, outcome-wide, for exposure measurement error than for outcome measurement error; with differential measurement error of the exposure or the outcome, the effect estimates themselves effectively constitute a bound for the strength of the differential measurement error needed to explain away the effects. For missing data, we recommend that, in most cases, this be handled outcome-wide, using multiple imputation, but that comparison be made with complete case analyses and that, in settings in which missing data is extensive or in which there are major discrepancies between complete case analyses and multiple imputation analyses, then the outcome-wide approach be abandoned and more detailed careful analyses be pursued taking into account the implications of the missing data for the analysis, separately for each outcome. Finally, we recommend also caution with the outcome-wide analytic approach when selection bias due to sample restriction is present as the biases in that setting can be substantial. Unmeasured confounding is always potentially present and can always be partially addressed with the E-value; measurement error is, in some sense, implicitly addressed by the effect estimates themselves; missing data and selection bias from restriction need to handled carefully in outcome-wide studies.

## 4. MULTIPLE TESTING METRICS

The outcome-wide analytic approach assesses the effect of a single exposure on numerous outcomes simultaneously. There might thus be concerns, in assessing numerous relationships, that there will be considerable potential for numerous false positives, where evidence seemingly arises for certain effects simply by chance, since so many different relationships are being evaluated. In this section we will discuss a variety of approaches to handle multiple testing. We comment on the use of Bonferroni correction as this remains a popular approach and in fact has various attractive properties not often appreciated. We suggest the reporting of other metrics as well related to methods that take into account the correlation among outcomes and that produce confidence intervals for the expected number of rejections that surpass a particular significance level threshold whilst taking into account correlations across outcomes.

### 4.1 Bonferroni Correction and Its Properties

The Bonferroni correction is perhaps still the most popular way of addressing issues of multiple testing (other than of course simply ignoring them, which is still arguably the most common). The Bonferroni correction is often motivated by preserving the type I error of the global null that all tested associations are in fact null. By dividing the nominal significance level of the test $\alpha$ (e.g. $\alpha=0.05$) by the number of tests, one is guaranteed, within a hypothesis testing framework, to reject the global null of no association at most $\alpha$ (e.g $\alpha=5\%$) of the time when the global null does in fact hold. While this is often the motivation presented for the Bonferroni correction, the correction itself does have a much stronger property. Suppose in an outcome-wide setting one were examining K exposure-outcome associations, and that, after Bonferroni correction, J associations were rejected at the α/K significance level. The standard property of the Bonferroni correction that is often pointed out is, as above, that no more than 5% of the time will one incorrectly conclude "There is at least one true association." But, with J rejections at the α/K significance level, one can in fact also consider the much stronger conclusion that "There are at least J true associations" and one will draw this conclusion, when it is false, at most 5% of the time (VanderWeele and Mathur, 2018). This is because even if there were in fact only J-1 true associations, the probability of rejecting J or more would still be less than [K-(J-1)] x α/K < K x α/K = α. The fact that this much stronger statement, like the rejection of the global null, also has only a 5% error rate gives the Bonferroni correction a much stronger interpretation when results surpass this more conservative threshold.

Such statements are also valid under any other procedure that strongly controls the familywise error rate (FWER), including those that are uniformly more powerful than the Bonferroni correction, such as the Holm (1979) procedure. It might therefore be tempting to conclude that whether one wants to make standard statements about the probability of at least one false positive, about the number of true associations as above, or both, the Bonferroni correction is obsolete and should be replaced with better FWER control procedures. However, this characterization is misleading because the Bonferroni correction in fact offers an even more stringent form of error control than do most FWER-control alternatives. Specifically, the Bonferroni correction controls the per-family error rate (PFER), which is the mean number of false positives divided by the number of tests (Gordon et al., 2007; Frane, 2015). To illustrate the distinction, suppose FWER is controlled via the uniformly more powerful Holm (1979) procedure. Then there is less than a 5% probability of obtaining at least one false positive, but if there is at least one false positive, there is no guarantee of how many there are; there could be one or 100. In contrast, the Bonferroni procedure guarantees that even if there is at least one false positive, there are still fewer than K x α in expectation. Other have argued persuasively that in many scientific contexts, every additional false positive is detrimental, and thus controlling the actual number of false positives (via PFER) is at least as important as controlling the presence or absence of any false positives (via FWER) (Frane, 2015). The Bonferroni correction may therefore be valuable in these contexts, even when one has also used more powerful FWER corrections. Thus, in spite of its conservative nature, we would recommend reporting the Bonferroni threshold in outcome-wide analyses, in addition to various other metrics described below.

While the Bonferroni is conservative and does not take into account correlation of the outcomes, it is often the case that, in settings in which sample sizes, are very large, such as many major cohort studies, and when only a moderate number of tests are being carried out, the Bonferroni correction will in fact often make relatively little difference in the magnitude of effect sizes that can generally be detected (VanderWeele and Mathur, 2018). Consider, for example, in a data analysis (Chen et al, 2018) related to what will be presented below with K=24 outcomes, sample size N=3,929 and with mean linear and logistic regression coefficient standard error of 0.031 across the various outcomes. In this context, for an outcome with standard error of 0.031, an effect estimate above 0.061 would suffice to pass the nominal α=0.05 significance level and an effect size above 0.095 would suffice to pass the Bonferroni-corrected significance level of α=0.05/24=0.0021. There is a relatively modest range of effect sizes, 0.061 to 0.095, for which the nominal significance level would be passed but the Bonferroni-corrected threshold would not be.  If variability of the outcomes were similar but with a sample size of N=10,000, an effect estimate above 0.038 (e.g. odds ratio of 1.039) would suffice to pass the nominal $\alpha=0.05$ significance level and an effect size above 0.060 (e.g. odds ratio of 1.062) would suffice to pass the Bonferroni-corrected significance level, of $\alpha=0.05/24=0.021$. Here the range of effect estimates for which the nominal significance threshold is passed but the Bonferroni corrected one is not, is even narrower, and arguably in many cases, that effect size range is sufficiently narrow to often not be of much scientific, policy, or public health importance (e.g. if the odds ratio is not even 1.062, the effect size may be too small to be of importance). Thus, with large sample sizes, in many settings, if the effect size estimate is sufficient to surpass the nominal threshold of $\alpha=0.05$ then it will very often be sufficient to pass the Bonferroni-corrected threshold as well.

Of course, just because the Bonferroni correction does not impose a severe penalty on the range of effect sizes that can be detected in some contexts, such as when the sample size is large and a moderate number of tests are being conducted, it does not follow from this that the penalty will always be negligible. In many settings, and perhaps especially in small- to medium- sized randomized trials, the sample sizes are often considerably smaller and the Bonferroni correction may constitute a much greater penalty for the relevant effect sizes that can be detected than is indicated here. This will also especially be the case in settings in which the study has been powered specifically to detect an effect for a primary outcome but in which many other secondary outcomes are examined as well. In such settings, or those with many outcomes, the Bonferroni correction might also likewise impose an especially severe penalty.

However, again in many outcome-wide studies, with large longitudinal cohorts especially, the penalty of the Bonferroni correction in terms of the potential effects sizes required to pass various thresholds is often very small and the added advantage of the strength of the conclusions that can be put forward might be considerable. One also need not definitively choose between using or not using the Bonferroni correction. Investigators can report the actual p-values themselves, and then also indicate the number of tests and what the Bonferroni corrected threshold would be. This allows the reader to assess evidence both as compared with the conventional nominal thresholds, and Bonferroni-corrected thresholds. In the section that follows we will also consider other useful multiple testing metrics as well.

**4.2 Additional Metrics Taking into Account Correlations**

We would recommend also reporting and commenting upon two other metrics that take into account correlation between outcomes in a single population. There are a variety of methods that have been proposed that preserve the familywise error rate (FWER), but are less conservative than the Bonferroni correction by taking into the account unknown correlations among the outcomes (e.g. Westfall and Young, 1993; Romano and Wolf, 2007). While it is difficult to provide definitive guidance on which of these various approaches will work best in any given setting, we believe the evidence from simulations currently points to very good performance of the approach put forward by Romano and Wolf (2007; cf. Mathur and VanderWeele, 2018), which can be used with parametric resampling approaches and generates datasets resembling the original data with the resampled test statistics then centered by their estimated values in the observed data in order to recover the null distribution. Thus in addition to the Bonferroni correction approach, when possible, it may be good to report the results of the Romano and Wolf (2007) resampling approach as well. Finally, it is, in addition, possible to report an interval with 95% coverage across repeated samples for the number of $\alpha$-level rejections that would be expected to occur under the global null of no association of the exposure on any of the outcomes while also taking into account the actual correlation structure among the outcomes themselves. We have developed theory to construct such a confidence interval and have developed an R package, NRejections, to implement this approach for continuous outcomes (Mathur and VanderWeele, 2018a); further theory will attempt to extend this to binary outcomes and logistic regression as well. A comparison of the actual number of $\alpha$-level rejections to the confidence interval can be informative as to the overall extent of the evidence for the presence and number of potential effects. Under certain technical conditions (that hold for example in linear regression models), the difference between the observed number of rejections and the upper limit of the 95% interval will constitute a lower bound on the number of true associations at least 95% of the time under repeated sampling. We think that this metric too can be informative.

Of course, none of these metrics is perfect, and the hypothesis testing framework is itself subject to many limitations and abuses (Rothman et al., 2008; Greenland et al., 2016). There is, moreover, nothing magical about the $\alpha=0.05$ threshold, or any other threshold (Benjamin et al., 2018), and these various approaches can be employed also across a range of significance level thresholds. However, reporting multiple of these measures that address multiple testing can help in that task of evidence synthesis and evaluation.

**4.3 Comment on Current Practices for Multiple Testing Correction**

While we believe that these various metrics, that take into account the fact that multiple associations are being examined in an outcome-wide study, are important, we do not think that the evidence from p-values that do not meet these multiple-testing-corrected thresholds should simply be ignored. The p-value is a continuous, not a dichotomous, metric. An extreme p-value of course does not guarantee that there is an actual association; nor does a large p-value guarantee that there is no association. The p-value is a continuous measure of evidence and should be treated as such. We believe it is still reasonable to comment upon the evidence for associations that do meet the nominal $\alpha$-level threshold, but do not meet this threshold after correction for multiple testing; and even reasonable to comment on effect sizes and possible evidence, or its absence, even for p-values above the nominal $\alpha$-level threshold. There is little difference in evidence between a p-value of 0.04 and 0.06. Moreover, ultimately, evidence is strongest when it is present in, and combined over, multiple studies. Meta-analysis provides one approach to such evidence synthesis and we believe that much of the strongest evidence in observational research comes from meta-analyses of numerous studies, and could be improved further by assessing their robustness to unmeasured confounding using meta-analytic analogues of the E-value (Mathur and VanderWeele, 2018b). The outcome wide analyses can provide input for such meta-analyses.

These considerations of not discarding evidence when it does not meet some multiple-testing-adjusted threshold are perhaps particularly relevant when one contrasts the outcome-wide approach with what is often current practice. Typically investigators, using the same data, will publish multiple papers of different exposure-outcome relationships, often including multiple papers using the same exposure. Much of current editorial practice allows comment upon associations that pass the nominal $\alpha=0.05$ threshold. However, it seems incongruous to allow comment upon such evidence if the same analyses are published over multiple papers versus within a single paper. The reporting of the actual continuous p-value and its comparison to different thresholds, both those with and without correction for multiple testing we do believe is worthwhile and helps the investigator and reader assess the overall evidence strength across the various outcomes. But no magical p=0.05 (with, or without, multiple-testing-adjustment) should be definitely imposed in discussing evidence and these considerations also need to be weighed within the context, and in light of the specific importance of avoiding false negatives (Rothman, 1990). We are in favor of reporting metrics related to multiple testing adjustment; we are not in favor of completely discarding evidence that does not surpass a given threshold; and again we believe that evidence will often only be particularly strong when it comes from more than one study, investigator, and population.

## 5. DATA ANALYSIS EXAMPLE

We will illustrate the outcome-wide approach with a data analysis concerning potential effects of parental warmth experienced in childhood on a variety of flourishing, mental health and health behavior outcomes. Following Chen et al. (2018a), we conducted longitudinal analyses of a subset of N = 2,984 subjects from the Midlife in the United States (MIDUS) cohort study, recruited to include siblings and twin pairs. For simplicity in these analyses, we randomly selected only one sibling from within each sibship (see Chen et al. (2018a) for the full analysis). In an initial wave of data collection (1995-1996), subjects recalled the parental warmth that they experienced during childhood as an average of separate scales of maternal and paternal warmth. In a second wave (2004-2006), the same subjects reported 13 continuous subscales of flourishing in emotional, psychological, and social domains, along with various mental health and health behavior outcomes. We assessed the association between a one-unit increase in standardized parental warmth (i.e., an increase of one standard deviation on the raw scale) with the standardized continuous composite flourishing score. We also examined potential effects on the 13 individual subscales treated separately and also the 3 standardized composite scores for each separate flourishing domain (emotional, psychological, and social). Other analyses assessed the associations between parental warmth and mental health problems (depression, anxiety) and adverse health behaviors and states (overweight/obesity, current or former smoking, heavy drinking, marijuana use, other substance use). All of our analyses controlled for age, sex, race, nativity status, parents' nativity status, number of siblings, whether the subject lived with biological parents, childhood socioeconomic status (SES), subjective SES, childhood welfare status, residential area, residential stability, maternal and paternal smoking, whether the subject lived with an alcoholic as a child, and religiosity. Multiple imputation was used to handle missing data (see supplemental methods for details: https://osf.io/tv3wu/). For continuous outcomes, we used ordinary least squares regression. For binary outcomes, we used Poisson regression if the sample prevalence was >10% (overweight/obesity, smoking, binge drinking, other substance use, and depression) and otherwise logistic regression (marijuana use and anxiety).

We expected correlation among the resulting 24 test statistics both because of conceptual similarities between the subscale variables (e.g., social acceptance and social integration) and because of the composite and domain measures' direct arithmetic relationships with the subscales. The 24 outcome measures had a median correlation magnitude of 0.25 (minimum = 0.0007; maximum = 0.88; 25th percentile = 0.08; 75th percentile = 0.43). For the composite flourishing outcome, controlling for demographics and childhood family factors, individuals reporting an additional standard deviation of parental warmth in childhood experienced greater mid-life flourishing by, on average 0.20 (95% CI: [0.16, 0.24]) standard deviations.

Table 2 reports the results of the outcome-wide analysis. Of the 24 outcomes considered individually, 18 were "significantly" associated with parental warmth at α = 0.05, 17 of which were also "significant" at α = 0.01. The directions of all 24 effects suggested that increased parental warmth was associated with improved flourishing outcomes. The E-values for these various associations and their confidence intervals are reported in Table 3. For a number of the flourishing outcomes, and also for depression, the E-value for the confidence interval is above 1.5, meaning that an unmeasured confounder that was associated with both high levels of parental warmth and with high levels of the outcome by risk ratios of 1.5-fold each, above and beyond the measured covariates could suffice to shift the confidence interval to the null but weaker confounding could not. The effect estimates on at least some of the flourishing outcomes thus seem reasonably robust to moderate amounts of unmeasured confounding. Various alternative codings of the exposure and the outcomes, motivated by the reporting considerations in the next section of the paper, that use tertiles of parental warmth, and that consider dichotomizations of the continuous outcomes are given in the Online Supplement in Tables S1-S4.

We now turn to the other multiple testing metrics. Under Bonferroni correction, 17 tests of all 24 remained "significant" ($\alpha \approx 0.002$). For the resampling-based measures, we had to restrict to the 17 continuous outcomes. Under the Romano and Wolf (2007) correction, 15 of the 17 tests of continuous outcomes remained "significant" at $\alpha = 0.05$ and 15 also at $\alpha = 0.01$. Using the methods described in Mathur and VanderWeele (2018a) to characterize the number of rejections, if parental warmth were in fact unassociated with all 17 continuous outcomes, we would expect 17 x 0.05 = 0.85 rejections at $\alpha = 0.05$ with a 95% null interval of [0, 5]; and 17 x 0.01 = 0.17 rejections at $\alpha = 0.01$ with a 95% null interval of [0, 2]. We thus observe 15-5=10 excess hits at $\alpha = 0.05$ and 15-2=13 excess hits at $\alpha = 0.01$ above what would be expected in 95% of samples under the global null. Overall, our outcome-wide analyses strongly support moderately sized effects of parental warmth on composite flourishing, as reported by Chen et al. (2018a). All data and code required to reproduce these analyses is publicly available and documented (https://osf.io/krjq2/).

## 6. REPORTING OF OUTCOME WIDE-ANALYSES

In this section we will discuss briefly convenient approaches to reporting results of outcome-wide analyses.

**6.1 Formatting of Tables**

A great deal of information is reported in the outcome-wide analyses being proposed here. An approach that we have found useful to report the considerable information in outcome-wide analyses in limited space is, in Table 1, to report on the demographics of the sample overall and/or across exposure groups (as is often done in practice). Because a single exposure is employed in outcome-wide analyses this will look analogous to what is already common practice in many empirical papers. Table 2 can report on the results from the primary outcome-wide analysis reporting on the magnitude of the association, its confidence interval, the p-value, with some indication of its surpassing or not various nominal and multiple-testing-corrected thresholds. For studies that report on both continuous and dichotomous outcomes, we have found it helpful, for reader presentation, to horizontally stagger the effect estimates so that risk ratios for binary outcomes are in one column, and regression coefficients for continuous outcomes are in another, as in Table 2 here. For continuous outcomes, both for the purposes of effect size comparison (which we will discuss further in Section 7.4 below) and to facilitate calculation of E-values, we recommend continuous outcomes in general be standardized to per-standard deviation changes, and perhaps especially so when the outcome scale is not well recognized (e.g. happiness or meaning scales). Table 3 can report on E-values for each outcome both for the estimate itself, and for the confidence interval. When standardized outcomes are used, it is important that the standard deviation of the outcome for the population be reported either in the paper or a supplement since such standard deviations can vary dramatically across populations depending on whether they are more homogeneous or diverse.

**6.2 Details of Measures in Online Supplements**

In our existing outcome-wide analyses (Chen et al., 2018ab; Chen and VanderWeele, 2018), we have often found it necessary to relegate some of the details on the measures used to an online supplement. Because the exposure is fixed in outcome-wide analysis and is the same for all outcomes, our recommendation is to discuss details of the exposure measurement in the text itself, and also to discuss issues related to the timing of the exposure, outcome, and covariates (the considerations in Sections 2.2-2.6 of this paper) in the body of the text also. However, when word counts are limited, as they often are with biomedical journals especially, we recommend placing more detailed descriptions of the measurement details and descriptive and psychometric properties of what are often an extensive number of outcomes and covariates in an online supplement. For some social science journals, with more generous word limits, this may not be necessary, but when words counts are limited, comment on variable timing and exposure measurement can be made in the text and covariate and outcome details can be placed in an online supplement.

**6.3 Effect Sizes for Continuous Exposures**

For continuous exposures, we recommend, for purposes of comparison, reporting primary analyses in one of three ways: (i) using the nominal exposure scale if this is well understood and selecting two values of the exposure that are substantively meaningful and comparing effects for them, or (ii) using a per-standard deviation standardized scale for the exposure if the scale used is not well understood; or (iii) dividing the exposure scale into tertiles or by median split and reporting effects sizes across the corresponding exposure categories. Reporting also need not be restricted to just one of these approaches and in general it may be desirable both to report on one of the approaches (i) or (ii) and also approach (iii). It can often be easier to explain to non-technical readers changes in outcomes across categories by referring to high and low levels of the exposure. Approaches (ii) and (iii) also make it easier to describe and present the results of E-value calculations since the exposure change is less arbitrary. Approach (ii) of using per-standard deviation changes is not always satisfactory, as a standard deviation change in the exposure will be relative to the population and might be in fact constitute a small change for a relatively homogeneous population, but a very large change for a diverse population. While standardized measures are problematic for effect size comparisons across populations (Greenland et al., 1986), they are less problematic within populations, provided the standard deviations themselves are also reported, since they then constitute only a rescaling of the original exposure scale. Examples of these additional analyses for the data example above are given in the Online Supplement to this paper.

**6.4 Effect Size Reporting and Conversions and Comparisons**

To facilitate comparison across effect sizes for different outcomes, continuous outcomes can also be converted to approximate risk ratios. We would not recommend this for the primary analyses but perhaps as an additional analysis for an online supplement. This can be done either by dichotomizing the continuous outcome at a substantively meaningful value or using a median split; or alternatively by the approximate conversion between standardized effect sizes and risk ratios referred to in Section 3.1 whereby a standardized effect sizes "d" with standard error $s_d$ is converted to an approximate risk ratio by RR≈exp(0.91×d) with approximate confidence interval (exp{0.91×d − 1.78×$s_d$}, exp{0.91×d + 1.78×$s_d$}). Again, this is derived using conversions often employed in the meta-analysis literature between common-outcome odds ratios and standardized effect sizes for continuous outcomes (Hasselblad and Hedges, 1995; Borenstein et al., 2009), and then between odds ratios and risk ratio (VanderWeele, 2017d). Examples of this for the analysis above are given in the Online Supplement.

## 7. ADVANTAGES OF OUTCOME-WIDE LONGITUDINAL DESIGNS

As noted in the introduction and as alluded to throughout the above text, carrying out outcome-wide longitudinal analyses has a number of advantages.

**7.1 Conveys More Information**

The most obvious advantage of the outcome-wide approach over individual studies of single exposure-outcome relationships is that far more information is conveyed in a single publication. The reader has a sense as to the effects of an exposure on a broad range of outcomes. There is an efficiency gain for the reader who need not search through countless studies; evidence for effects of the exposure on numerous outcomes is presented at once. There is also an efficiency gain for the researcher, and for the research community. The effort to go through the peer-review process for a large number of distinct papers, each reporting a single exposure-outcome association is considerable; it is considerable for the researcher, and it is considerable for the editors and peer reviewers. If a study design is strong for one exposure-outcome relationship, it will also often, though not always, be strong for numerous other outcomes as well. We believe knowledge will advance more rapidly if the outcome-wide approach were broadly adopted. The number of total publications would go down, but in an era wherein this number has grown exponentially, this reduction would arguably be no bad thing. It might be argued that this lower number could be problematic for the researcher for promotion purposes. Our view is that such decisions should be made principally on the underlying substantive contribution of research, rather than simply the number of publications. Moreover, while an outcome-wide analysis is certainly more work than the analysis of a single exposure-outcome relationship, once the principles and reporting practices are mastered, it is not dramatically more work; and given the much greater contribution to the literature we believe that the slightly lower number of publications will often be offset by the greater prominence and contribution of the studies themselves. We believe that ultimately the outcome-wide approach will be of benefit both to the broad research community and our knowledge base, and also to the individual researchers themselves.

The conveying of dramatically more information in an outcome-wide analysis is also arguably of benefit for policy and for public health (VanderWeele, 2017a). For exposures such as hormone replacement therapy, or moderate alcohol consumption, which may have beneficial effects on some outcomes and harmful effects on others it will be desirable to see all of these at once in making informed public health and policy recommendation. Ideally one would arguably want the effects of the exposures on numerous flourishing outcomes, broadly conceived (VanderWeele, 2017b). The neglect of this can lead to papers and results that are arguably of little relevance. A recent paper reported positively on the beneficial effects of divorce for weight loss (Kutob et al, 2017). We think that the association is plausible due to the desire to re-enter the dating market. However, given the well-established negative effects of divorce on so many other outcomes (Marks and Lambert, 1998; Waite and Gallagher, 2000; Wilcox, 2011; Shor et al., 2012), the effect on weight loss is almost beside the point. An outcome-wide approach that examined numerous outcomes would put the weight loss result into proper context. Again, from a policy and public health perspective, we believe that it will be often be best to examine effects on numerous outcomes simultaneously.

There are of course also limitations inherent in the attempt to assess the effect of an exposure on numerous effects simultaneously. The type and extent of theoretical discussion that often accompanies empirical analyses in the social sciences of a single exposure-outcome relationship will not be possible for each and every outcome in an outcome-wide analysis. Broad theory might, however, be put forward with regard to why the exposure should affect numerous, rather than one, outcome. Outcome-wide longitudinal analyses might also be viewed as principal input for subsequent theorizing. The relationship between theory and empirical work is bi-directional (King et al., 1994), and certain effects, if detected empirically, may give rise to new theory, even if their initial discovery was not theoretically motivated.

**7.2 Reporting of Null Results**

It has been frequently noted one problematic aspect of current practices in scientific publishing is that it is difficult to publish null results (Rosenthal, 1979; Ziliak and McCloskey, 2008). Many journals do not want to publish research that simply says there is no effect. However, it has been argued that in some cases null results can be as or more important or informative than results suggesting evidence for an effect (Abadie, 2018). An outcome-wide analysis allows for the reporting of null results, along with those for which there seems evidence for an effect, in a single paper. We believe that this too would be an important contribution of the outcome-wide approach for more easily allowing for the publication of null results.

**7.3 Less Temptation to Choose Models**

Another advantage of the outcome-wide approach is that it may lead to fewer instances in which the analysis results are substantially biased by investigator choice after looking at the data. We believe there will be less temptation, when employing the approach described above for outcome-wide analyses, to choose among different models and different sets of covariates to obtain the results the investigator desires. While this should not be done even for a single model, there is inevitable temptation to make decisions on analysis retrospectively, after seeing the results, and selecting those most similar to what one hopes to find. This phenomenon is sometimes referred to as one of "researcher degrees of freedom" (Simmons et al., 2016) or a "garden of forking paths in the analysis of data" (Gelman and Loken, 2014). The outcome-wide approach does not eliminate this danger entirely. It is still possible to run numerous outcome-wide analyses, each outcome-wide analysis with a different set of covariates, or with a different type of modeling approach and select among them. However, if, within any given outcome-wide analysis, each outcome in that specific outcome-wide uses the same covariates, and the same modeling approach, then the "researcher degrees of freedom" will be dramatically reduced as compared with if all of these same choices were able to be made separately, and differently, for each and every outcome. Said another way, it will be more difficult to "optimally choose" results across numerous outcomes in accord with investigator expectations when the investigator is constrained to make similar modeling choices across the outcomes under consideration. We believe that this too is an advantage of the outcome-wide approach.

It could, however, be argued that with outcome-wide analyses there will still be temptation to examine numerous outcomes and then only selectively report the results of some of these. This certainly is a danger. We hope that the previous comment on the opportunity to more easily report null results will in part mitigate this danger. Indeed the reporting of null results may even, in fact, provide some evidence that the positive results obtained are not due solely to unmeasured confounding, if some of the outcomes might plausibly serve as negative controls (Lipsitch et al., 2010). The question of the selection of outcomes is indeed an important one. When data are available and effects on human well-being are of interest, we would recommend selecting several outcomes, perhaps the broadest possible, from each of the aforementioned flourishing domains (VanderWeele, 2017b): happiness and life satisfaction, mental and physical health, meaning and purpose, character and virtue, close social relationship, and financial security. Of course, in most datasets there will be richer data on certain of these outcomes than on others. Pre-registration of analytic plans can also mitigate some of the dangers of researcher degrees of freedom; however, with existing secondary data, it can sometimes be difficult to be certain whether the registration preceded or followed preliminary analyses.

**7.4 The Comparison of Effect Sizes**

Another advantage of the outcome-wide approach is the capacity to compare effect sizes of the exposure across outcomes. Is the effect of parental warmth on autonomy or on life satisfaction greater? If these associations are reported in different studies using different populations it can be very difficult to make these determinations. A difference in the magnitude of association may be due to larger effects on some outcomes than on others, but could also be due to the fact that different populations are used in different studies; age or race or income differences across the populations may be responsible for the differing effects sizes on two different outcomes assessed in two different studies. An outcome-wide analysis allows, at least for the sample under consideration, a more clear and direct comparison of effect sizes. Effects may of course still differ across populations and results from one study should not necessarily be generalized to other populations, but as outcome-wide studies are undertaken in different populations for the same sets of exposure-outcome relationships, it may become clearer on which outcomes effects of an exposures are particularly large across populations.

# 8. DESIGN VARIATIONS

In this section we will consider variations on, and alternatives to, the outcome-wide longitudinal design and how analogues of it might, or might not, be applied in other contexts or with other approaches intended to assess causal effects.

## 8.1 Challenges of Exposure-Wide Designs

There has been a recent suggestion that the research community begin to move towards "exposure-wide" studies, in which associations between an outcome and many exposures – possibly very many exposures – are assessed simultaneously (Ioannidis, 2016). This has perhaps arisen in part because of the success of genome-wide association studies. However, as argued elsewhere (VanderWeele, 2017a), due to the nature of confounding, attempts at "exposure-wide epidemiologic" studies are likely to be plagued by biases, in contrast to the "outcome-wide" approach laid out above. The notion of an exposure-wide epidemiologic study is that a researcher could select a specific outcome, regress it upon a wide range of different exposures, either one-at-a-time or all simultaneously, assess which relationships are most substantial, and for which there is the strongest statistical evidence of an association, and, provided appropriate control is made for multiple testing, thereby potentially gain insight into the underlying causes of the disease or outcome under study. This approach has effectively been what has been used in genome-wide association studies, and these have now yielded thousands of replicated associations between genetic variants and various diseases (Hunter, 2012; Welter et al., 2014).

The difference between genetic exposures and almost all others, and the difference that creates problems for an exposure-wide analyses, lies in the nature of confounding. In a genome wide association study, although hundreds of thousands of variants are examined, it is often thought to be the case that, subject to control for population stratification (often done say by principal components analysis adjustment strategies), the association between the variant and the outcome is roughly unconfounded (Hunter, 2012). While a particular variant may serve as a proxy for the true effect of another, it is the case that once the genome is fixed, each variant is acting on the outcome, possibly in conjunction with, but not by altering the value of, any other variant. This is manifestly not the case with environmental, behavioral, and social exposures, wherein one exposure is likely to affect many others downstream. Each exposure will thus likely require a distinct set of other variables to control for confounding, with the confounding variables for a particular exposure consisting only of other exposures that are temporally prior to it. If we include all of our exposures in the model and some of these are downstream from others, then the downstream exposures will likely mediate, and potentially block, the effects of prior exposure.

This creates two potential problems. First, for each exposure, the association estimate will, at best, represent the direct effect of the exposure not through any of the other exposures in the model downstream of it. We are not getting the overall total effect of each exposures. If there are numerous subsequent exposures that mediate the effect of the prior exposure then the importance of the prior exposure (in terms of its overall influence on the outcome) might be severely misrepresented as noted above in Section 2.3. Second, it is now well documented in the methodological literature that if control is made for mediating variables on pathways from exposure to outcome, then any unmeasured common cause of the mediating variable and the outcome can induce bias; spurious associations between exposure and outcome can be generated even if the exposure has no effect on the outcome whatsoever. This problem is sometimes referred to in the literature as one of "collider stratification bias" (Cole et al., 2010; Hernán and Robins, 2018). When considering multiple exposures simultaneously the likelihood of such biases is substantial. In an exposure-wide study, the number of potential instances of such biases that must be considered when dozens, or hundreds, of exposures are considered simultaneously, is mind-boggling, when each exposure must have a separate set of confounders. Empirical studies currently struggle with these issues in studies of a single exposure. It is arguably not reasonable then to think that we could do this adequately when numerous exposures are considered at once. Moreover, even if we could, we would still only be obtaining direct effects as above.

As discussed in Section 2, if the total effect of the exposure on the outcome is desired, then adjustment should not be made for variables that might be affected by the exposure. The implications of this, as indicated above, is that for each individual exposure, we will likely need a distinct set of confounding variables. We cannot make the decision about confounding for all variables at once when we are supposedly examining the effects of multiple exposures. A single regression model will not suffice; nor will simply looking at each bivariate association one at a time, as in genome-wide studies. This arguably creates difficulties for a simple approach to exposure-wide studies. In contrast, with an outcome-wide study, while some variables may confound the relationship between the exposure and one outcome, but not another, we do still have the option, unlike in the exposure-wide epidemiologic setting, of simply controlling for all, or almost all, of the variables prior to the exposure as described in Section 2 above. We have the option of attempting to make confounding control decisions for all outcomes at once. Said another way, with the exposure fixed, the set of all variables temporally prior to the exposure stays the same even when we change the outcome. With the outcome fixed, the set of variables temporally prior to the exposure changes as we change the exposure.

## 8.2 Lagged Exposure-Wide Designs

However, an alternative exposure-wide approach with a restricted set of exposures, that are roughly contemporaneous with one another, may turn out to be more feasible (VanderWeele, 2017a). With cohort data for which repeated measures of exposures are available, one might examine a single outcome at the end of follow up (call this wave $W_3$) and fit a series of regressions, each of which controls for all exposures simultaneously in one wave ($W_1$) but then also includes a single subsequent exposure – one per regression – from the next wave ($W_2$). We might refer to this as a "lagged exposure-wide design." An approach such as this would still make all confounding control decisions simultaneously (all covariates and exposures available at $W_1$) in all regressions and thus could be automated. As per discussions above about covariate timing, one would want $W_1$ and $W_2$ to not temporally be too far apart so as to risk the possibility of substantial time-dependent confounding.

With a single outcome Y at wave 3, and exposures $(A_1,…,A_J)$ at wave 2, and a set of covariates C that ideally includes all of the same exposures at wave 1, and also demographic and other covariates we could fit a series of regression models:

$E[Y|a_j,c] = \alpha_j+\beta_j a_j+\gamma_j'c$

and likewise for other regression models that may be of interest. For each exposure $A_j$ at wave 2, provided the confounding control assumption holds that $Y(a_j) \perp A_j | C$, the coefficient $\beta_j$ in each model will provide a consistent estimate of the causal effect of exposure $A_j$ on outcome Y on the

relevant scale corresponding to the regression model being used. For a linear regression model $E[Y|a_j,c] = \alpha_j+\beta_j a_j+\gamma_j{}'c$, we have that, provided $Y(a_j) \perp A_j | C$, then $\beta_j = E[Y(A_j=1) - Y(A_j=0)|c]$.

Such analyses would not give a complete picture of all of the exposures relevant for the outcome since they are effectively restricted to those measured at a given point in time, thus precluding e.g. relevant childhood exposures if the primary waves of the analysis ($W_1$ and $W_2$) were in adulthood. Such analyses are also effectively restricted to exposures that can change over time within the relevant time interval (i.e. between $W_1$ and $W_2$). However this lagged exposure-wide approach might still be useful for gaining insight into the determinants of an outcome at a particular point in time. In this regard, they are arguably also useful from a policy perspective in determining what can, or cannot, effectively change the outcome of interest at that time.

Of course the outcome-wide approach could itself be employed across numerous exposures giving something of a hybrid between the outcome-wide and exposure-wide approaches. See Betancourt et al. (2015) for such an example in examining the effects, for former child soldiers in Sierra Leonne, of schooling, community acceptance, stigma, and other exposures, on numerous subsequent outcomes.

### 8.3 Interaction Outcome-Wide Studies

The outcome-wide approach we have discussed has concerned a single exposure, but, if we employed such an approach with two exposures, we could also assess potential interaction between the two exposures across the different outcomes of interest. If the two exposures, which we will denote here by A and X, are relatively contemporaneous then this could be done in a relatively straightforward way within a regression context by fitting a series of models of the form:

$$E[Y_k|a,x,c] = \alpha_k+\beta_k a+\delta_k x+\phi_k x+\gamma_k{}'c$$

or likewise for other regression models that may be of interest. One could report the main effects and the interactions of both of the exposures A and X outcome-wide. One could also potentially report the proportion of the effect due to just the first exposure alone, due to just the second exposure alone, and due to their interaction (VanderWeele and Tchetgen Tchetgen, 2014). Such measures to assess the proportion attributable to interaction can also, across models and outcome types, all be converted to a difference scale for comparative purposes (VanderWeele and Tchetgen Tchetgen, 2014).

If the exposures are not contemporaneous but rather one affects the other and there are potential intermediate confounders that are affected by the first exposure and then go on to confound the relationship between the second exposure and the outcome, then the confounding control assumptions become more complex (VanderWeele, 2009; Robins et al., 2000). The approach could still be employed in principle but causal models, such as marginal structural models (Robins et al., 2000), that extend beyond the simple regression approach described above, would need to be employed. The same is also true for causal effects of time-varying exposures to which we now turn.

### 8.4 Outcome-Wide Studies for Causal Effects of Time-Varying Exposures

As noted in Section 2, with exposures like exercise, or employment, or religious service attendance, that change over time, one can attempt to assess the causal effects of an entire trajectory of the exposures. The confounding control assumptions required for this, and the causal modeling approaches needed to do this are then more complex and beyond the scope of the paper. Good introductions to causal inference with time-varying exposures are given elsewhere (Robins, 1992, Robins et al., 2000; Robins and Hernán, 2009; Hernán and Robins, 2018) and the reader is referred there for further discussion.

However, as regards an outcome-wide approach, this could in principle be done also with causal effects of a time-varying exposure, and similar principles to what was described above in Section 2 would arguably be applicable but extended to the time-varying exposure. In general we believe that this will typically be more feasible for marginal structural models (Robins et al., 2000) and parametric g-formula approaches (Garcia-Aymerich et al., 2014; Hernán and Robins, 2018) than for structural nested models (Robins, 1992; Robins and Hernán, 2009). With marginal structural models and parametric g-formula approaches the same models could potentially be employed across outcomes and only the final outcome under consideration would perhaps need to be changed. With structural nested model approaches because many of the statistical estimation options require numeric grid searches derived from the outcomes themselves, this could be more challenging, and involved, in an outcome-wide setting. But once again, there is nothing in principle that would prohibit carrying out an outcome-wide analysis for the causal effects of a time-varying exposure.

**8.5 Quasi-Experimental Outcome-Wide Designs**

The outcome-wide approach could also in principle be applied in various quasi-experimental designs. The reasonableness of this may vary by context. When an instrumental variable analysis is being used to assess causal effects, the outcome-wide approach may be reasonably plausible when the instrument for the treatment or exposure is itself randomized, as may be the case when assignment to treatment is taken as an instrument for treatment compliance or when the draft lottery number is used as an instrument for participation in the army. One could assess, say, the local average treatment effects (Angrist et al., 1996) across numerous different outcomes. However, in contexts in which the instrument is not subject to some degree of randomization and careful substantive arguments need to be made to justify the exclusion restriction, then an outcome-wide approach will likely be less plausible as these exclusion restriction arguments would have to be made for each and every outcome.

For regression discontinuity designs (Lee and Limieux, 2010; Bor et al., 2014), if the running variable is such that the rule for treatment assignment is deterministic, or at least follows a definitive randomized protocol, an outcome-wide approach could potentially be employed. One could assess the local conditional treatment effect across numerous different outcomes. If, however, substantive arguments are needed to justify that no other change relevant to the outcome occurs when the running variable reaches the discontinuity threshold and these arguments need to be made for each and every outcome, then the outcome-wide approach may be less reasonable in such contexts.

With interrupted time-series designs (Morgan and Winship, 2015; Bernal et al., 2017) it may be more difficult to carry out outcome-wide as careful assessment of the outcome trajectories, before and after the intervention, would be required for each and every outcome.

## 8.6 Mediator-Wide Studies

Another variation on the outcome-wide or exposure-wide design would be within the context of mediation, and to then consider a mediator-wide design wherein both the exposure and the outcome are fixed but numerous potential mediators are examined one at a time. This is potentially problematic because if the mediators affect one another but are evaluated as mediators singly, one at a time, then this can generate considerable biases (VanderWeele and Vansteelandt, 2013; VanderWeele, 2015). The approach may only be plausible if the mediators themselves are measured relatively contemporaneously, shortly after the exposure, and then have relatively little effect on one another over the relevant time horizon (VanderWeele, 2015). See Kim and VanderWeele (2018) for an example of a mediator-wide study assessing potential mediators for the effect of religious service attendance on all-cause mortality. In settings in which the mediators do affect one another it may be more reasonable to assess the effect mediated through the entire set of mediators considered jointly (VanderWeele and Vansteelandt, 2013), rather than one at a time.

## 9. CONCLUSION

In this paper we have put forward a new template for empirical studies intended to assess causal effects across outcomes: the outcome-wide longitudinal design. We have discussed principles of confounding control in these designs, metrics to assess unmeasured confounding, and additional metrics to deal with questions of multiple testing. Much of the paper has provided or referenced theoretical justification for the proposed approach, but some of the material that has been discussed has been more in the spirit of tentative guidelines for the approach. We have been employing this approach in many of our own recent analyses and have referenced some of these examples (Chen et al., 2018ab; Chen and VanderWeele, 2018; Betancourt et al., 2015), but guidelines will perhaps be refined as more analyses are carried out. The paper has laid out a vision for the types of analyses that might be possible – a new template. The material discussed is not so much a theory of causal inference – though we have discussed a number of theoretical contributions that have motivated the approach – but rather it is a theory of causal inference for addressing a particular set of questions. It is theory for an approach to causal inference that attempts to assess the effects of a single exposure at a single period of time on numerous subsequent outcomes. Numerous other questions within causal inference such as regards time-varying exposures (Robins, 1992, Robins et al., 2000; Robins and Hernán, 2009; Hernán and Robins, 2018), mediation analysis (Imai et al., 2010; VanderWeele, 2015), censoring by death (Hayden et al., 2005; Rubin, 2006), contagion and interference (Sobel, 2006; Hudgens and Halloran, 2008; Tchetgen Tchetgen and VanderWeele, 2012), and local treatment effect (Angrist et al., 1996) will require other approaches and other theory. The causal inference theory laid out here is thus, in some ways, somewhat narrow in scope.

On the other hand, we believe that the outcome-wide longitudinal design has the potential to become the norm for a particular set of causal questions intended to assess causal effects on numerous outcomes using longitudinal or panel data and confounding control. We believe it has the potential to largely replace studies that currently assess only a single exposure-outcome relation using regression models or propensity scores. There will, of course, always be need for careful evaluation of single exposure-outcome relationships. But in many contexts, when many outcomes are of interest and relevance, as we believe they often are, then the outcome-wide approach will, we think, often be preferable. Of course the value, and even possibility, of such outcome-wide studies depends critically on having a broad range of outcomes available and, to that end, we strongly encourage data collection aspects on numerous aspects of human flourishing broadly construed (VanderWeele, 2017b). In such contexts, we believe that use of outcome-wide designs will help the field with more objective inference, with the reporting of null results, with more consistent evaluation of potential unmeasured confounding, with the comparison of effect sizes, and with better assessment of policy and public health relevance. These advantages will thereby also contribute to a more rapid and accurate advancement of knowledge and, if a broad range of outcomes are examined, with the promotion of human flourishing. We encourage therefore the use of this design in practice and look forward to future refinements and developments.

## REPRODUCIBILITY

All code required to reproduce the applied example is publicly available (https://osf.io/krjq2/). The dataset used for the applied example is publicly available through the Inter-University Consortium for Political and Social Research (ICPSR); we detail how to access the dataset and reproduce the applied example in our public repository (https://osf.io/tdcyw/).

**Table 1. Covariates for confounding control in outcome-wide analyses**

| Domain | Covariate |
|---|---|
| Demographic | Race |
| | Age |
| | Gender |
| | Marital Status |
| Economic, Social and Political | Income |
| | Education |
| | Employment |
| | Social integration |
| | Neighborhood |
| | Religious service attendance |
| | Political affiliation |
| Health | Self-rated health |
| | Number of health conditions |
| | Exercise |
| | Smoking |
| | Alcohol Consumption |
| | Depression |
| Psychological | Happiness |
| | Loneliness |
| | Parental warmth |
| | Purpose/Meaning |
| | Big Five Personality |

**Table 2. Longitudinal associations of parental warmth (1994-1995) with health and well-being outcomes (2004-2006).**

| Health and well-being outcome | B | OR or RR | 95% CI | *p*-value[a] | Romano Correction |
|---|---|---|---|---|---|
| **Overall composite** | | | | | |
| Overall flourishing (continuous) | 0.20 | | [0.16, 0.24] | <0.0001*** | * |
| **Flourishing domain composites** | | | | | |
| Emotional well-being | 0.19 | | [0.16, 0.23] | <0.0001*** | * |
| Social well-being | 0.13 | | [0.09, 0.16] | <0.0001*** | * |
| Psychological well-being | 0.18 | | [0.14, 0.22] | <0.0001*** | * |
| **Flourishing subscales** | | | | | |
| Emotional Well-Being | | | | | |
|   Positive affect | 0.18 | | [0.14, 0.22] | <0.0001*** | * |
|   Life satisfaction | 0.16 | | [0.12, 0.20] | <0.0001*** | * |
| Social Well-Being | | | | | |
|   Meaningfulness of society | 0.04 | | [-0.00, 0.08] | 0.053 | |
|   Social integration | 0.15 | | [0.11, 0.19] | <0.0001*** | * |
|   Social acceptance | 0.09 | | [0.05, 0.13] | <0.0001*** | * |
|   Social contribution | 0.08 | | [0.04, 0.12] | <0.0001*** | * |
|   Social actualization | 0.07 | | [0.03, 0.11] | 0.0005*** | * |
| Psychological Well-Being | | | | | |
|   Autonomy | 0.07 | | [0.03, 0.11] | 0.0004*** | * |
|   Environmental mastery | 0.13 | | [0.09, 0.17] | <0.0001*** | * |
|   Personal growth | 0.09 | | [0.05, 0.13] | <0.0001*** | * |
|   Positive relations | 0.23 | | [0.19, 0.26] | <0.0001*** | * |
|   Purpose in life | 0.04 | | [-0.00, 0.07] | 0.083 | |
|   Self-acceptance | 0.19 | | [0.15, 0.23] | <0.0001*** | * |
| **Adverse health behaviors** | | | | | |
| Overweight or obese | | 0.99 | [0.95, 1.05] | 0.823 | N/A |
| Smoking | | 0.95 | [0.90, 1.00] | 0.052 | N/A |
| Binge drinking | | 0.98 | [0.87, 1.10] | 0.726 | N/A |
| Marijuana use | | 0.81 | [0.65, 1.00] | 0.053 | N/A |
| Any other drug use | | 0.85 | [0.75, 0.95] | 0.006** | N/A |
| **Mental health problems** | | | | | |
| Depression | | 0.77 | [0.69, 0.86] | <0.0001*** | N/A |
| Anxiety | | 0.76 | [0.58, 1.00] | 0.047* | N/A |

Abbreviations: CI = confidence interval; OR = odds ratio; RR = risk ratio.
n =2,948 for all analyses. Estimates are from ordinary least squares, Poisson, or logistic regression on multiply-imputed datasets and are adjusted for age, sex, race, nativity status, parents' nativity status, number of siblings, whether the subject lived with biological parents, childhood socioeconomic status (SES), subjective SES, childhood welfare status, residential area, residential stability, maternal and paternal smoking, whether the subject lived with an alcoholic as a child, and religiosity. For binary outcomes, we used Poisson regression if the sample prevalence was >10% (overweight/obesity, smoking, binge drinking, other substance use, and depression) and otherwise logistic regression (marijuana use and anxiety).
a: * = $p < 0.05$; ** = $p < 0.01$; *** = significant under Bonferroni correction, counting all outcome measures ($p < 0.002$).
b: This correction could be applied only to the continuous outcomes, so corrected only for multiplicity among those 17 hypothesis tests. N/A indicates a non-continuous outcome.

**Table 3. Robustness to unmeasured confounding (E-values[a]) for causal effects of parental warmth (1994-1995) on health and well-being outcomes (2004-2006).**

| Health and well-being outcome | E-value for point estimate | E-value for CI |
|---|---|---|
| **Overall composite** | | |
| Overall flourishing (continuous) | 1.69 | 1.59 |
| | | |
| **Flourishing domain composites** | | |
| Emotional well-being | 1.67 | 1.57 |
| Social well-being | 1.49 | 1.38 |
| Psychological well-being | 1.64 | 1.53 |
| | | |
| **Flourishing subscales** | | |
| Emotional Well-Being | | |
|   Positive affect | 1.64 | 1.53 |
|   Life satisfaction | 1.59 | 1.48 |
| Social Well-Being | | |
|   Meaningfulness of society | 1.23 | 1.00 |
|   Social integration | 1.56 | 1.46 |
|   Social acceptance | 1.39 | 1.26 |
|   Social contribution | 1.37 | 1.25 |
|   Social actualization | 1.34 | 1.20 |
| Psychological Well-Being | | |
|   Autonomy | 1.34 | 1.20 |
|   Environmental mastery | 1.50 | 1.39 |
|   Personal growth | 1.39 | 1.27 |
|   Positive relations | 1.76 | 1.66 |
|   Purpose in life | 1.22 | 1.00 |
|   Self-acceptance | 1.66 | 1.56 |
| | | |
| **Adverse health behaviors** | | |
| Overweight or obese | 1.08 | 1.00 |
| Smoking | 1.30 | 1.00 |
| Binge drinking | 1.17 | 1.00 |
| Marijuana use | 1.46 | 1.00 |
| Any other drug use | 1.64 | 1.27 |
| | | |
| **Mental health problems** | | |
| Depression | 1.92 | 1.59 |
| Anxiety | 1.56 | 1.04 |

*Abbreviations: CI = confidence interval.*
[a] *: See VanderWeele and Ding (2017) for the formula for calculating E-values.*

**Figure. 1. Diagram illustrating how control for prior exposure ($A^{prior}$) can further reduce potential for unmeasured confounding (U)**

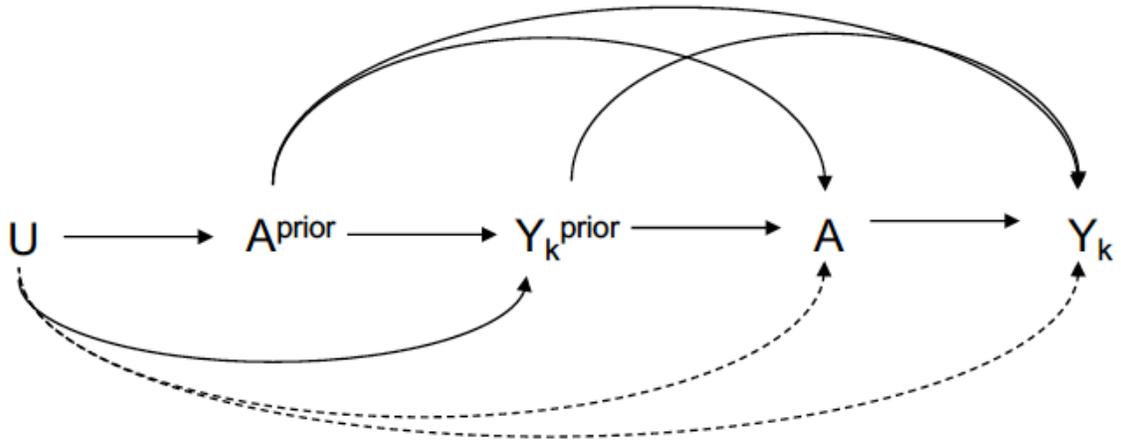

**Table S1. Complete-case analysis on the longitudinal association of parental warmth (standardized score) with health and well-being outcomes in mid-life (The Midlife in the United States Study 1994-1995 to 2004-2006 wave, N ranged from 2,618 to 2,742[a])**

| Health and well-being outcomes | B[b] | RR[c] | 95% CI | P value |
|---|---|---|---|---|
| **Overall composite** | | | | |
| Overall flourishing | 0.21 | | 0.17, 0.25 | <.0001*** |
| **Flourishing domain composites** | | | | |
| Emotional well-being | 0.20 | | 0.16, 0.24 | <.0001*** |
| Social well-being | 0.12 | | 0.08, 0.16 | <.0001*** |
| Psychological well-being | 0.19 | | 0.15, 0.23 | <.0001*** |
| **Flourishing subscales** | | | | |
| Emotional well-being | | | | |
|     Positive affect | 0.18 | | 0.14, 0.22 | <.0001*** |
|     Life satisfaction | 0.17 | | 0.13, 0.21 | <.0001*** |
| Social well-being | | | | |
|     Meaningfulness of society | 0.04 | | -0.004, 0.08 | .075 |
|     Social integration | 0.14 | | 0.10, 0.18 | <.0001*** |
|     Social acceptance | 0.09 | | 0.05, 0.13 | <.0001*** |
|     Social contribution | 0.08 | | 0.04, 0.12 | <.0001*** |
|     Social actualization | 0.07 | | 0.03, 0.11 | .0009*** |
| Psychological well-being | | | | |
|     Autonomy | 0.08 | | 0.04, 0.12 | .0003*** |
|     Environmental mastery | 0.13 | | 0.09, 0.17 | <.0001*** |
|     Personal growth | 0.10 | | 0.06, 0.14 | <.0001*** |
|     Positive relations | 0.24 | | 0.20, 0.28 | <.0001*** |
|     Purpose in life | 0.04 | | -0.01, 0.08 | .092 |
|     Self-acceptance | 0.19 | | 0.15, 0.23 | <.0001*** |
| **Adverse health behaviors** | | | | |
| Overweight or obese | | 1.00 | 0.97, 1.03 | .87 |
| Current or former smoking | | 0.95 | 0.91, 0.99 | .0089** |
| Heavy drinking | | 1.00 | 0.90, 1.11 | .99 |
| Marijuana use | | 0.79 | 0.64, 0.98 | .033* |
| Any other drug use | | 0.83 | 0.72, 0.95 | .0064** |
| **Mental Health** | | | | |
| Depression | | 0.78 | 0.70, 0.87 | <.0001*** |
| Anxiety | | 0.62 | 0.46, 0.83 | .0016*** |

Abbreviations: B= standardized beta; RR=risk ratio; CI=confidence interval.
[a] The full analytic sample was restricted to those who had valid data on parental warmth and all covariates. The actual sample size for each analysis varied depending on the number of missing values for each outcome under investigation. All models controlled for age, sex, race, nativity status, parents' nativity status, number of siblings, and other childhood family factors (childhood family structure, family SES, area of residence, residential stability, parents' smoking status, alcoholics in the family and family religiousness).
[b] All continuous outcomes were standardized (mean=0, standard deviation=1), and B was the standardized effect size.
[c] Effect estimates for the outcomes of marijuana use, any other drug use and anxiety were odds ratios; these outcomes were rare (prevalence <10%), so the odds ratios would approximate the RRs. Effect estimates for other dichotomized outcomes were RRs.
*p<0.05 before Bonferroni correction; **p<0.01 before Bonferroni correction; ***p<0.05 after Bonferroni correction (the p value cutoff for Bonferroni correction =0.05/24 outcomes=0.002)

**Table S2.  The longitudinal association of parental warmth tertiles with health and well-being outcomes in mid-life (The Midlife in the United States Study 1994-1995 to 2004-2006 wave, N=2,948 [a])**

| Health and well-being outcomes | Parental warmth | | | | | | | |
|---|---|---|---|---|---|---|---|---|
| | *Middle tertile vs. bottom tertile* | | | | *Top tertile vs. bottom tertile* | | | |
| | $B$ [b] | $RR$ [c] | 95% CI | P value | $B$ [b] | $RR$ [c] | 95% CI | P value |
| **Overall composite** | | | | | | | | |
| Overall flourishing | 0.23 | | 0.15, 0.32 | <.0001*** | 0.43 | | 0.33, 0.52 | <.0001*** |
| **Flourishing domain composites** | | | | | | | | |
| Emotional well-being | 0.16 | | 0.07, 0.25 | .0004*** | 0.40 | | 0.31, 0.50 | <.0001*** |
| Social well-being | 0.22 | | 0.13, 0.31 | <.0001*** | 0.25 | | 0.16, 0.35 | <.0001*** |
| Psychological well-being | 0.20 | | 0.11, 0.29 | <.0001*** | 0.40 | | 0.30, 0.50 | <.0001*** |
| **Flourishing subscales** | | | | | | | | |
| Emotional well-being | | | | | | | | |
|    Positive affect | 0.16 | | 0.07, 0.25 | .0004*** | 0.39 | | 0.29, 0.48 | <.0001*** |
|    Life satisfaction | 0.12 | | 0.03, 0.21 | .008** | 0.33 | | 0.23, 0.42 | <.0001*** |
| Social well-being | | | | | | | | |
|    Meaningfulness of society | 0.09 | | 0.00, 0.18 | .040* | 0.10 | | 0.01, 0.20 | .037* |
|    Social integration | 0.25 | | 0.16, 0.33 | <.0001*** | 0.32 | | 0.22, 0.41 | <.0001*** |
|    Social acceptance | 0.15 | | 0.06, 0.24 | .001*** | 0.17 | | 0.07, 0.26 | .0007*** |
|    Social contribution | 0.11 | | 0.02, 0.19 | .017* | 0.16 | | 0.07, 0.26 | .0007** |
|    Social actualization | 0.17 | | 0.08, 0.26 | .0002*** | 0.14 | | 0.04, 0.24 | .0056** |
| Psychological well-being | | | | | | | | |
|    Autonomy | 0.10 | | 0.01, 0.19 | .037* | 0.20 | | 0.10, 0.30 | <.0001*** |
|    Environmental mastery | 0.15 | | 0.06, 0.24 | .0011*** | 0.27 | | 0.18, 0.37 | <.0001*** |
|    Personal growth | 0.08 | | -0.01, 0.17 | .100 | 0.20 | | 0.10, 0.29 | <.0001*** |
|    Positive relations | 0.23 | | 0.14, 0.32 | <.0001*** | 0.48 | | 0.38, 0.57 | <.0001*** |
|    Purpose in life | 0.06 | | -0.03, 0.15 | .18 | 0.07 | | -0.02, 0.17 | .13 |
|    Self-acceptance | 0.22 | | 0.13, 0.30 | <.0001*** | 0.41 | | 0.32, 0.51 | <.0001*** |
| **Adverse health behaviors** | | | | | | | | |
| Overweight or obese | | 1.01 | 0.91, 1.13 | .81 | | 1.00 | 0.88, 1.13 | .99 |
| Current or former smoking | | 0.89 | 0.79, 1.01 | .073 | | 0.89 | 0.78, 1.02 | .091 |
| Heavy drinking | | 0.92 | 0.71, 1.18 | .51 | | 0.93 | 0.71, 1.23 | .62 |
| Marijuana use | | 0.69 | 0.44, 1.10 | .12 | | 0.47 | 0.26, 0.83 | .0089** |
| Any other drug use | | 0.69 | 0.53, 0.89 | .0051** | | 0.62 | 0.46, 0.84 | .0021** |
| **Mental health problems** | | | | | | | | |
| Depression | | 0.69 | 0.53, 0.89 | .0051** | | 0.62 | 0.46, 0.84 | .0021** |
| Anxiety | | 0.59 | 0.31, 1.13 | .11 | | 0.64 | 0.31, 1.30 | .21 |

Abbreviations: B= standardized beta; RR=risk ratio; CI=confidence interval.
[a] Multiple imputation was performed to impute for missing data on the exposure, outcomes and covariates. All models controlled for age, sex, race, nativity status, parents' nativity status, number of siblings, and other childhood family factors (childhood family structure, family SES, area of residence, residential stability, parents' smoking status, alcoholics in the family and family religiousness).
[b] All continuous outcomes were standardized (mean=0, standard deviation=1), and B was the standardized effect size.
[c] Effect estimates for the outcomes of marijuana use and anxiety were odds ratios; these outcomes were rare (prevalence <10%), so the odds ratios would approximate the RRs. Effect estimates for other dichotomized outcomes were RRs.
*p<0.05 before Bonferroni correction; **p<0.01 before Bonferroni correction; ***p<0.05 after Bonferroni correction (the p value cutoff for Bonferroni correction =0.05/24 outcomes=0.002)

**Table S3. The longitudinal association between parental warmth tertiles and flourishing in mid-life, with the effect estimates for continuous outcomes converted to the risk ratio scale (The Midlife in the United States Study 1994-1995 to 2004-2006 questionnaire wave, N=2,948 [a])**

|  | Parental warmth | | | | | |
|---|---|---|---|---|---|---|
|  | *Middle tertile vs. bottom tertile* | | | *Top tertile vs. bottom tertile* | | |
| Health and well-being outcomes | RR [b] | 95% CI | P value | RR [b] | 95% CI | P value |
| **Overall composite** | | | | | | |
| Overall flourishing | 1.24 | 1.14, 1.34 | <.0001*** | 1.47 | 1.35, 1.61 | <.0001*** |
| **Flourishing domain composites** | | | | | | |
| Emotional well-being | 1.16 | 1.07, 1.25 | .0004*** | 1.44 | 1.32, 1.57 | <.0001*** |
| Social well-being | 1.22 | 1.13, 1.32 | <.0001*** | 1.26 | 1.16, 1.37 | <.0001*** |
| Psychological well-being | 1.20 | 1.11, 1.30 | <.0001*** | 1.44 | 1.32, 1.57 | <.0001*** |
| **Flourishing subscales** | | | | | | |
| Emotional well-being | | | | | | |
|    Positive affect | 1.16 | 1.07, 1.25 | .0004*** | 1.42 | 1.30, 1.55 | <.0001*** |
|    Life satisfaction | 1.12 | 1.03, 1.21 | .008** | 1.34 | 1.23, 1.47 | <.0001*** |
| Social well-being | | | | | | |
|    Meaningfulness of society | 1.09 | 1.00, 1.18 | .040* | 1.10 | 1.01, 1.19 | .037* |
|    Social integration | 1.25 | 1.15, 1.36 | <.0001*** | 1.33 | 1.23, 1.45 | <.0001*** |
|    Social acceptance | 1.15 | 1.06, 1.24 | .001*** | 1.16 | 1.07, 1.27 | .0007*** |
|    Social contribution | 1.10 | 1.02, 1.19 | .017* | 1.16 | 1.07, 1.27 | .0007** |
|    Social actualization | 1.17 | 1.08, 1.27 | .0002*** | 1.13 | 1.04, 1.24 | .0056** |
| Psychological well-being | | | | | | |
|    Autonomy | 1.09 | 1.01, 1.18 | .037* | 1.20 | 1.10, 1.31 | <.0001*** |
|    Environmental mastery | 1.28 | 1.18, 1.40 | .0011*** | 1.28 | 1.18, 1.40 | <.0001*** |
|    Personal growth | 1.07 | 0.99, 1.16 | .100 | 1.20 | 1.10, 1.31 | <.0001*** |
|    Positive relations | 1.23 | 1.13, 1.33 | <.0001*** | 1.55 | 1.42, 1.69 | <.0001*** |
|    Purpose in life | 1.06 | 0.97, 1.15 | .18 | 1.07 | 0.98, 1.17 | .13 |
|    Self-acceptance | 1.22 | 1.12, 1.32 | <.0001*** | 1.46 | 1.33, 1.59 | <.0001*** |

Note: RR=risk ratio; CI=confidence interval.

[a] Multiple imputation was performed to impute for missing data on the exposure, outcomes and covariates. All models controlled for age, sex, race, nativity status, parents' nativity status, number of siblings, and other childhood family factors (childhood family structure, family SES, area of residence, residential stability, parents' smoking status, alcoholics in the family and family religiousness).

[b] Linear regression models were first used to estimate the standardized beta B (where the outcome follows a normal distribution, all continuous outcomes were standardized at mean=0, standard deviation=1). The effect estimates were then converted to the risk ratio scale by applying the approximation RR≈exp (0.91*B), and CI≈ (exp{0.91×B-1.78×sB}, exp{0.91×d+1.78×sB}).

*p<0.05 before Bonferroni correction; **p<0.01 before Bonferroni correction; ***p<0.05 after Bonferroni correction (the p value cutoff for Bonferroni correction =0.05/24 outcomes=0.002)

**Table S4. The longitudinal association between parental warmth tertiles and flourishing in mid-life, with all continuous outcomes dichotomized at the median split (The Midlife in the United States Study 1994-1995 to 2004-2006 questionnaire wave, N=2,948[a])**

| | Parental warmth | | | | | |
| --- | --- | --- | --- | --- | --- | --- |
| | *Middle tertile vs. bottom tertile* | | | *Top tertile vs. bottom tertile* | | |
| Health and well-being outcomes | RR[b] | 95% CI | P value | RR[b] | 95% CI | P value |
| **Overall composite** | | | | | | |
| Overall flourishing | 1.22 | 1.07, 1.40 | .0039** | 1.45 | 1.26, 1.66 | <.0001*** |
| **Flourishing domain composites** | | | | | | |
| Emotional well-being | 1.16 | 1.01, 1.33 | .030* | 1.38 | 1.20, 1.59 | <.0001*** |
| Social well-being | 1.14 | 1.00, 1.29 | .057 | 1.19 | 1.04, 1.37 | .014* |
| Psychological well-being | 1.18 | 1.03, 1.35 | .016* | 1.42 | 1.23, 1.63 | <.0001*** |
| **Flourishing subscales** | | | | | | |
| Emotional well-being | | | | | | |
|    Positive affect | 1.14 | 0.99, 1.30 | .069 | 1.37 | 1.19, 1.57 | <.0001*** |
|    Life satisfaction | 1.14 | 0.97, 1.34 | .099 | 1.45 | 1.23, 1.70 | <.0001*** |
| Social well-being | | | | | | |
|    Meaningfulness of society | 1.09 | 0.96, 1.23 | .19 | 1.03 | 0.90, 1.17 | .68 |
|    Social integration | 1.23 | 1.08, 1.39 | .0019*** | 1.31 | 1.15, 1.51 | <.0001*** |
|    Social acceptance | 1.17 | 1.02, 1.35 | .025* | 1.23 | 1.06, 1.42 | .0066** |
|    Social contribution | 1.06 | 0.93, 1.20 | .38 | 1.13 | 0.99, 1.30 | .07 |
|    Social actualization | 1.12 | 0.98, 1.28 | .083 | 1.16 | 1.01, 1.33 | .037* |
| Psychological well-being | | | | | | |
|    Autonomy | 1.06 | 0.93, 1.21 | .36 | 1.16 | 1.01, 1.33 | .038* |
|    Environmental mastery | 1.13 | 0.98, 1.30 | .089 | 1.28 | 1.10, 1.47 | .001*** |
|    Personal growth | 1.06 | 0.93, 1.21 | .36 | 1.19 | 1.04, 1.37 | .0096** |
|    Positive relations | 1.27 | 1.11, 1.46 | .0005*** | 1.62 | 1.41, 1.86 | <.0001*** |
|    Purpose in life | 1.05 | 0.92, 1.19 | .50 | 1.07 | 0.92, 1.23 | .38 |
|    Self-acceptance | 1.18 | 1.03, 1.36 | .019* | 1.43 | 1.24, 1.65 | <.0001*** |

Note: RR=risk ratio; CI=confidence interval.

[a] Multiple imputation was performed to impute for missing data on the exposure, outcomes and covariates.
[b] All continuous outcomes were dichotomized at the median split. Regression models were used to estimate RR (Poisson distribution). All models controlled for age, sex, race, nativity status, parents' nativity status, number of siblings, and other childhood family factors (childhood family structure, family SES, area of residence, residential stability, parents' smoking status, alcoholics in the family and family religiousness).
*p<0.05 before Bonferroni correction; **p<0.01 before Bonferroni correction; ***p<0.05 after Bonferroni correction (the p value cutoff for Bonferroni correction =0.05/24 outcomes=0.002)